\documentclass[twocolumn,prd,showpacs,floatfix]{revtex4}
\usepackage{graphicx}
\usepackage{dcolumn}
\usepackage{bm}
\begin{document}
\title{Harmonic oscillator force between heavy quarks}
\author{ Stanis{\l}aw D. G{\l}azek } 
\affiliation{ Institute of Theoretical Physics, Warsaw University, 
              ul.  Ho{\.z}a 69, 00-681 Warsaw, Poland }
\date{\today}
\begin{abstract}
A renormalization group procedure for effective particles is applied to quantum 
chromodynamics of one flavor of quarks with large mass $m$ in order to calculate 
light-front Hamiltonians for heavy quarkonia, $H_\lambda$, using perturbative 
expansion in the coupling constant $\alpha_\lambda$. $\lambda$ is the renormalization 
group parameter with the interpretation of an inverse of the spatial size of the 
color charge distribution in the effective quarks and gluons. The eigenvalue 
equation for $H_\lambda$ couples quark-anti-quark states with sectors of a larger 
number of constituents. The coupling to states with more than one effective gluon, 
and interactions in the quark-anti-quark-gluon sector, are removed at the price of 
introducing an ansatz for the gluon mass, $\mu^2$. The simplified equation is used 
to evaluate a new Hamiltonian of order $\alpha_\lambda$ that acts only in the 
effective quark-anti-quark sector and in the non-relativistic limit turns out to 
contain the Coulomb term with Breit-Fermi corrections and spin-independent harmonic 
oscillator term with frequency $\omega = [(4/3)(\alpha_\lambda/\pi)]^{1/2}\lambda 
(\lambda/m)^2(\pi/1152)^{1/4}$. The latter originates from the hole excavated in 
the overlapping quark self-interaction gluon clouds by the exchange of effective 
gluons between the quarks. The new term is largely independent of the details of 
$\mu^2$ and in principle can fit into the ball park of phenomenology. The first 
approximation can be improved by including more terms in $H_\lambda$ and solving 
the eigenvalue equations numerically.
\end{abstract}
\maketitle

\section{ Introduction }
\label{sec:I}

The purpose of this article is to describe a procedure that starts from 
a quantum chromodynamics (QCD) with only one flavor of massive quarks 
and produces the Schr\"odinger equation for heavy quarkonia in a single 
formulation of the theory. Only the first approximation for the final 
Hamiltonian is evaluated. In this simplest version, the procedure involves 
a guess for the gluon mass term. But the guess appears to have little 
influence on the result. The procedure itself is not limited to the 
simplest version and the gluon mass ansatz can be tested in future in 
refined calculations and phenomenology. The procedure is relativistic and 
can be used for quarkonia in arbitrary motion, which is a pre-requisite 
for application in high-energy processes. Chiral symmetry is explicitly 
broken in the case of heavy quarks and the issue of the spontaneous 
breaking of the symmetry is ignored. 

The approach described here stems from the similarity renormalization 
group procedure for Hamiltonians \cite{similarity}, which has been 
applied to QCD \cite{long} using the light-front (LF) form of dynamics 
\cite{front}. A new ingredient is the boost-invariant creation and 
annihilation operator calculus for effective quarks and gluons (see
below). Otherwise, the LF approach is known for a long time, mainly as 
a candidate for connecting two qualitatively different models of hadrons: 
the parton model in the infinite momentum frame (IMF) \cite{partons, 
frontreview1} and the constituent model in the rest frame of a hadron 
\cite{PDG}. Many contributions in that area \cite{frontreview2} have 
followed the seminal work on exclusive processes \cite{LB}. Through the 
boost invariance and precisely defined notion of effective constituents, 
the approach described here aims at providing a bridge between the two 
models of hadrons in a single theoretical framework. The case of a heavy 
quarkonium is chosen here as the simplest one to begin with and test the 
method.

In the LF dynamics, the evolution of states is traced along the direction 
of a light-like four-vector $n^\mu$, for which $n^2=0$. With the conventional 
choice of $n = (1,0,0,-1)$, the variable $nx = x^0 + x^3 \equiv x^+$ plays 
the role of time while $P^- = P^0 - P^3$ is the Hamiltonian. In order to 
define it for bare particles in QCD, one has to choose a gauge. No serious 
alternative exists to $nA=A^+=0$. But the equation $D_\mu F^{\mu +} = 
j^+$ implies a constraint that is analogous to the Gauss law and forces 
the Fourier components of $A^-$ to contain inverse powers of the kinematical 
momentum $k^+$. Since $k^+$ ranges from 0 to infinity, the inverse powers 
of $k^+$ produce singularities in the region around zero.
 
One can impose a lower bound on $k^+$, such as $k^+ > \delta^+$, to 
regulate the theory \cite{p+1, p+2, long, DLCQ}. The parameter $\delta^+$ 
becomes a smallest unit of momentum that any particle, physical or 
virtual, can carry in such discretized theories. But the fixed unit 
breaks the boost invariance required for connection between the IMF 
and the rest frame of any hadron. Namely, when some physically relevant 
$P^+$ is made large, the smallest allowed $x = \delta^+/P^+$ becomes 
small. In the IMF $P^+ \rightarrow \infty$ and the same small-$x$ 
divergences re-appear despite the presence of $\delta^+$. The key singularity
is related to the $dx/x$ distribution of gluons in the parton model and 
seems to require a dynamical mechanism to remove. One cannot just vary 
$\delta^+$ together with $P^+$ because boosts cannot change the cutoff 
in a quantum theory constructed {\it ab initio} \cite{Wigner}. 

There exists a possibility that the small-$k^+$ singularities are related 
to the properties of the vacuum state. The sum rules for heavy quarkonia 
\cite{condensates1} include quark and gluon condensates \cite{condensates2} 
that may participate in the dynamics in the small-$k^+$ region \cite{frontreview1, 
long, condensates3, csb1, csb2, csb3, z, RozowskiThorn}. In the QCD picture 
with such nontrivial ground state \cite{th1, cdg}, and relativistic bound-state 
excitations of this state in the form of $\pi$-mesons, one can hardly hope 
to resolve the small-$k^+$ singularity easily. The situation simplifies 
a lot in the case of quarks with mass $m \gg \Lambda_{QCD}$. The small-$x$ 
singularity and a non-perturbative binding mechanism for quarks and gluons 
can interplay with each other without interference from the vacuum. In the 
case of light quarks, a similar interplay may be at work and contribute to 
the saturation mechanism of partons \cite{lip, mue1, mue2, mue3}, but that 
issue is not addressed here.

The gluon mass ansatz is introduced to represent effects of the non-abelian 
interactions. The ansatz is inserted at the level of solving the eigenvalue 
equation for the Hamiltonian $H_\lambda$, where $\lambda$ is the renormalization 
group (RG) parameter. The procedure of introducing the mass ansatz is similar 
to the one proposed in \cite{long} and later discussed in simplified matrix 
models \cite{kolleg, modelafbs}. New elements are the limitation of power 
counting to the relative-motion variables, the exact boost invariance, removal 
of small-$x$ divergences through the mass ansatz as a function of the relative 
momenta, and no need for ad hoc potentials in the first approximation. All 
these features will be described in detail later. 

In brief, $H_\lambda = T_\lambda  + V_\lambda$, where $T_\lambda$ is the 
kinetic energy operator made of all terms that are bilinear in the creation 
and annihilation operators for the effective particles, and $V_\lambda$ 
represents all other terms. The parameter $\lambda$ defines the width of 
momentum-space form factors in $V_\lambda$. For some value of $\lambda =
\lambda_0 \sim 1 $ GeV, one can freely add to $H_{\lambda_0}$ a term of 
the form $[1 - (\alpha_{\lambda_0} /\alpha_s)^n]\mu^2$, where $n \geq 2$. 
The number 2 ensures that corrections to the first approximation occur 
first in the fourth-order of expansion of $H_{\lambda_0}$ in powers 
of $g_0 = g_{\lambda_0}$, $\alpha_0 = g_0^2/(4\pi)$. This is the lowest 
order at which a perturbative shift in the gluon energy in any state 
can influence the contribution of that state to the dynamics of any other 
state in perturbation theory, keeping intact the QED-like small coupling 
expansion scheme with a Coulomb potential. $\mu^2$ stands for the mass term 
that one assigns to the effective gluons of the transverse size $1/\lambda_0$. 
The interpretation of $1/\lambda$ as the spatial size of the color charge 
distribution in the corresponding effective particles is based on the 
feature mentioned above that $V_\lambda$ contains vertex form factors of 
width $\lambda$ in momentum space. The mass ansatz contributes to the 
invariant masses through $\mu^2/x$, where $x$ is the fraction of the 
longitudinal momentum carried by the gluon. $\alpha_s$ denotes the large, 
relativistic value of the coupling constant in QCD at the scale $\lambda_0$
with a true $\Lambda_{QCD}$ in this scheme. The term with $\mu^2$ vanishes 
for $\alpha_0 = \alpha_s$. Nevertheless, only $\mu^2$ counts when the ratio 
$\alpha_0/\alpha_s$ is small. Thus, in the weak coupling expansion,
\begin{eqnarray} 
\label{ansatz1}
H_{\lambda_0} = T_{\lambda_0} + \mu^2 + [V_{\lambda_0} - (\alpha_0/\alpha_s)^n \mu^2] 
\end{eqnarray}
and the term $[V_{\lambda_0} - (\alpha_0/\alpha_s)^n \mu^2]$ is
treated as a source of small corrections in comparison to $\mu^2$. 

The requirement of cancellation of the small-$x$ singularities in the 
effective dynamics imposes some perturbatively determined constraints 
on the otherwise non-perturbative ansatz for $\mu^2$. These constraints 
restrict the behavior of $\mu^2$ as a function of the gluon motion 
with respect to other constituents. In future, the refined versions 
of the same procedure may provide constraints that come closer to the 
actual behavior of gluons. This behavior is hoped to be uncovered in 
computer simulations that one may build around the first approximation. 
Then, the extrapolation to $\alpha_0 = \alpha_s$ can recover the original 
theory from a few terms in the weak-coupling expansion if $\mu^2$ 
approximates the behavior of effective gluons well. Initially, the 
gluon mass term is viewed as a function of the relative momenta and 
$\Lambda_{QCD}$. The latter depends on $\alpha_0$ as $\lambda_0 \exp
{(-c/\alpha_0)}$ with a positive constant $c$. This means that 
$\Lambda_{QCD}$ vanishes to all orders in the perturbative expansion, 
and $\mu^2$ is considered to be on the order of 1. Before one knows more, 
$\mu$ can only be estimated on the basis of implications for the resulting 
Schr\"odinger equation. The size of $\mu$ can be compared with four scales: 
$\Lambda_{QCD}$, $\lambda_0$, $m$, and the Bohr momentum scale, $k_B = 
\alpha_0 m/2$, which is distinguished non-perturbatively by the Coulomb 
interaction. But if $\mu^2$ is right, then $[V - (\alpha_0/\alpha_s)^n 
\mu^2]$ must be a source of only small corrections in the whole range of 
couplings between 0 and $\alpha_s$. This is taken for granted in the 
present article. Since the first approximation turns out to be not sensitive 
to the details of $\mu$, new information can be obtained only in the 
refined calculations.

The value of the weak coupling expansion scheme for Hamiltonians is 
that it starts from a local theory and leads to $H_\lambda$ that is 
capable of describing physically relevant non-perturbative dynamics 
even if $H_\lambda$ is calculated only in low orders. This idea is 
known to work in the case of QED: the Coulomb potential accounts for 
highly non-perturbative dynamics of atoms, including the nature of 
chemical bond, while the Hamiltonian itself is only of the formal order 
of $\alpha$. Condensed matter physics illustrates this point in still
wider domain. 

But when one applies the weak coupling expansion idea to QCD \cite
{long}, one faces the fact that the strong coupling constant rises to 
10, 30, or even 100 times larger value than in the case of atoms or 
positronium in QED. This  leads to a complex interplay between the 
perturbative and non-perturbative parts of the calculation, enhanced 
dependence of observables on the RG parameter $\lambda$, problems with 
obtaining the Poincare symmetry in solutions, and amplification of 
artifacts due to the small-$k^+$ regularization. Most of the problems 
seem to come from perturbation theory in the RG part of the calculation. 
An exact RG procedure by definition provides $H_\lambda$ whose structure 
depends on $\lambda$ but the spectra and $S$-matrix elements do not. 
However, when one uses expansion in powers of $\alpha_0$ and then 
extrapolates to $\alpha_0 = \alpha_s$, a considerable dependence of the 
eigensolutions on $\lambda$ can ensue because of missing many terms. This 
is visible in models that are asymptotically free and produce bound states 
\cite{modelafbs, optimization}. One has to find the right value for 
$\alpha_s$ at given $\lambda_0$ from fits to bound-state observables, 
and perform consistency checks for whole sets of different observables 
\cite{optimization1}. In QCD, such checks involve the unknown functions 
of momenta in the finite parts of the ultraviolet counterterms, unknown 
terms depending on $\Lambda_{QCD}$, and artifacts of the regularization 
of small-$k^+$ divergences. So many unknowns suggest the possibility 
that the approach may never achieve the desired level of predictive 
power. But the simplicity of the harmonic potential found here in the 
first approximation illustrates that there is a high degree of order 
in the rich structure of $H_\lambda$. It is a consequence of preserving 
all seven kinematical symmetries of the LF scheme in the RG procedure. 
These symmetries limit the large number of terms that are allowed by the 
LF power counting using absolute momentum variables \cite{long}, to a 
much smaller number of terms that depend only on the relative momenta of 
the constituents. The LF symmetries must also be respected by the initial 
regularization prescription for the required counterterms to be simple.

The renormalization group procedure for effective particles (RGEP), which 
is employed here, regulates the ultra-violet and small-$k^+$ divergences 
by vertex factors $r$ that are inserted only in the interaction terms. In 
the case of the small-$k^+$ singularities, these factors limit only the 
ratios of momenta $k^+$. The ratios are limited from below by a dimensionless
parameter $\delta$. Details of the factors $r_\delta$ are explained later. 
Every creation or annihilation operator, labeled by momentum $k^+$, enters 
the initial interaction Hamiltonian together with a corresponding factor 
$r_\delta(x)$, where $x=k^+/p^+$ and $p^+$ is the sum of all momenta that 
label all creation operators, or, equivalently, all annihilation operators
in the same interaction term (see Appendix \ref{A:H}). In QCD with 
$r_\delta(x) \sim x^\delta$, $H_\lambda$ contains the coupling constant 
$g_\lambda$, which depends on the scale $\lambda$ in the same asymptotically 
free way \cite{g_QCD} that characterizes the running coupling constant 
dependence on the renormalization scale in the Feynman diagrams \cite{af1, af2}. 

Hamiltonians $H_{\lambda_0}$ with small $\lambda_0$ are worth studying 
because their eigenstates can be expanded in the effective particle basis 
in the Fock space and the wave functions in this expansion are expected 
to correspond to the constituent picture of hadrons. This is envisioned 
in analogy to the models based on Yukawa theory \cite{largep}. The 
interactions are suppressed by the form factors and cannot copiously create 
new constituents, even if the coupling constant $\alpha_0$ becomes large. 
Exotic hadron states may have their probability distributions shifted 
in the number of effective particles above the constituent quark model 
values of 2 or 3 \cite{exotics}. The effective dynamics can be in agreement 
with requirements of special relativity even if it is limited to a small 
number of effective constituents, and the RGEP provides rules for 
constructing the representation of the Poincare group \cite{algebra}. 
But the key feature is that the transition between the bare and effective 
degrees of freedom is made in one and the same formalism. There is no need 
to match different formulations of the theory, such as in the case of lattice 
theory and the continuum perturbation theory in the Minkowski space \cite
{match1, match2, match3}, with none of the parts meant to cover the whole 
range of relevant scales on its own. 

One more comment is required concerning the small-$x$ divergences in
the eigenvalue equation for $H_\lambda$. In the initial studies that 
used $\delta^+$ to limit bare particles' momenta and employed coupling 
coherence to derive certain $H_\lambda$ \cite{gcoh1, gcoh2, gcoh3, QQgcoh, 
QQ}, one could keep only a quark-anti-quark sector in the corresponding 
eigenvalue problem and the resulting equation was finite in the limit 
$\delta^+ \rightarrow 0$. Similarly, no small-$k^+$ divergences were 
encountered in the case of gluonium approximated by states of only two 
gluons \cite{gluonium}. In contrast, the present RGEP approach demands
inclusion of states that contain an additional effective gluon which is
needed to cancel small-$x$ divergences. For example, if one keeps only 
a pair of the effective quark and anti-quark, the leading non-relativistic 
(NR) terms are free from the small-$x$ singularities \cite{RGEP} but 
relativistic corrections are singular \cite{divx1, divx2}. When the 
additional gluon is included, the condition of cancellation of the small-$x$ 
divergences becomes a guide in understanding the gluon dynamics. The rules 
of including the gluons must be brought under quantitative control and the 
well-known case of heavy quarkonia provides a laboratory for testing the 
approach based on the gluon mass ansatz. The tests require a first approximation 
to begin with and a candidate is identified in the next sections.

This paper is organized as follows. Section \ref{sec:H} describes the 
initial Hamiltonian of LF QCD with one heavy flavor and the procedure for 
deriving an effective $H_\lambda$. Section \ref{sec:E} discusses the 
eigenvalue equation for a quarkonium, and introduces the ansatz for the 
gluon mass term. Small-$x$ effects in the dynamics are described in 
Section \ref{sec:sx}. The resulting potential in the Schr\"odinger 
equation for a $Q\bar Q$ bound state is described in Section \ref{sec:SE}. 
Section \ref{sec:C} provides a brief summary and outlook. Appendices 
contain key details required for completeness.
  
\section{Hamiltonians}
\label{sec:H}
The regularized canonical Hamiltonian of LF QCD with one heavy flavor 
of quarks, $H$, is given in Appendix \ref{A:H}. It includes ultraviolet 
counterterms. This section describes the main features of $H$ and the RGEP 
derivation of the effective Hamiltonian $H_\lambda$ with a finite width 
$\lambda$. $H_\lambda$ is independent of the ultraviolet regularization 
factors $r_\Delta$ in $H$ when $\Delta \rightarrow \infty$. The small-$x$ 
regularization factors $r_\delta$, which are also present in $H$, and their 
role in $H_\lambda$, will be discussed later. 

The initial Hamiltonian has the structure
\begin{equation}
\label{H}
H = H_{\psi^2} + H_{A^2} + H_{\psi A \psi} + H_{(\psi\psi)^2} + X
\end{equation} 
where the term $H_{\psi^2}$ denotes the kinetic energy operator 
for quarks, $H_{A^2}$ the kinetic energy operator for gluons, 
$H_{\psi A \psi}$ is the interaction term that couples gluons to 
quarks, $H_{(\psi\psi)^2}$ is the instantaneous interaction between 
quarks, and $X$ denotes all other terms including the counterterms.

In terms of the creation operators for bare particles, $b^\dagger$ 
for quarks, $d^\dagger$ for anti-quarks, $a^\dagger$ for gluons, and 
the corresponding annihilation operators, the kinetic energy terms 
are of the form,
\begin{equation}
\label{tquark}
H_{\psi^2} = \sum_{\sigma c} \int [k] {k^{\perp \, 2} + m^2 \over k^+}
  \left[b^\dagger_{k\sigma c}b_{k\sigma c} + 
  d^\dagger_{k\sigma c }d_{k\sigma c} \right] \, ,
\end{equation}
and
\begin{equation}
\label{tgluon}
H_{A^2} = \sum_{\sigma c} \int [k] {k^{\perp \, 2} \over k^+} \,
  a^\dagger_{k\sigma c}a_{k\sigma c} \, ,
\end{equation}
where $k$ denotes the three kinematical momentum components, $k^+$ and 
$k^\perp = (k^1,k^2)$. The symbol in a bracket, such as $[k]$, refers 
to the integration measure, 
\begin{eqnarray}
[k] = {dk^+ d^2 k^\perp \over 16 \pi^3 k^+} \, .
\end{eqnarray}
The subscript $c$ stands for color and $\sigma$ for spin. The mass $m$ 
is assumed to be very large in comparison to $\Lambda_{QCD}$. 

The quark-gluon coupling terms in $H_{\psi A \psi}$ that 
preserve the number of quarks and anti-quarks, have the form
\begin{equation}
\label{Y}
Y = \, g \sum_{123} \, \int[123]  \, \tilde r_{3,1} 
\left[j_{23} \, b^\dagger_2 a^\dagger_1 b_3 -
 \bar j_{23} \, d^\dagger_2 a^\dagger_1 d_3 + h.c. \right] \, .
\end{equation} 
The regularization factor $\tilde r_{3,1}$ is singled out
to indicate its presence. The coefficients $j_{23}$ and $\bar 
j_{23}$, are functions of the quark and gluon colors, spins, 
and momenta, with all details provided in the Appendix \ref{A:H}.
These coefficients contain the three-momentum conservation 
$\delta$-function factors, denoted by $\tilde \delta$, color 
factors $t^1_{23}$, and products of spinors, $j^\mu_{23} = 
\bar u_2 \gamma^\mu u_3$ and $\bar j^\mu_{32} = \bar v_3 
\gamma^\mu v_2$. The latter are contracted with polarization 
vectors for gluons, so that
\begin{equation}
j_{23} = \tilde \delta \, t^1_{23} \, 
g_{\mu \nu} \, j^\mu_{23} \,  \varepsilon^{\nu \, *}_1 \, ,
\end{equation}
and 
\begin{equation}
\bar j_{23} = \tilde \delta \, t^1_{32} \, 
g_{\mu \nu} \, \bar j^\mu_{32} \, \varepsilon^{\nu \, *}_1 \, .
\end{equation}
The instantaneous term $H_{(\psi\psi)^2}$ contains 
\begin{eqnarray}
Z & = & -g^2 \,\sum_{1234}\int[1234] \,\tilde\delta \, t^a_{12} t^a_{43}
\left[ j^+_{12} \bar j^+_{34}/ (k^+_1-k^+_2)^2\right]
\nonumber \\
& \times & 
 [\tilde r_{1,2} \tilde r_{4,3}+ \tilde r_{2,1} \tilde r_{3,4}]\, 
  b_1^\dagger d_3^\dagger d_4 b_2 \, . 
\end{eqnarray}

The current factors $j$ and the gluon polarization vectors grow
with the relative transverse momenta of the interacting particles, 
$\kappa^\perp$. These can increase to infinity and the regularization 
factors $r_\Delta$ are introduced to limit the range to a finite one. 
In addition, there are small-$x$ divergences due to the inverse powers 
of $x$, especially in the gluon polarization vectors that contain 
terms proportional to $\kappa^\perp/x$. The small-$x$ singularities 
are regulated by factors $r_\delta$. 

The RGEP procedure generates ultraviolet counterterms contained in the 
operator $X$ in Eq. (\ref{H}) and renders the effective particle 
Hamiltonian $H_\lambda$ which is independent of $r_\Delta$. The procedure 
is defined order by order in the formal expansion in powers of the bare 
coupling constant $g$. This expansion is eventually re-written in terms 
of the effective coupling constant, $g_\lambda$, which replaces $g$ in 
$H_\lambda$ and depends on the ratio $\lambda/\Lambda_{QCD}$ \cite{g_QCD}. 
The procedure is designed so that energy differences in denominators of 
the perturbative evaluation of $H_\lambda$ are limited from below by 
$\lambda$. Therefore, no infrared divergences are encountered in the 
evaluation of $H_\lambda$. Also, no perturbative intrusion into the 
binding mechanism is generated when $\lambda$ is kept above the scale 
of typical relative momenta of the bound-state constituents. These 
features qualify the RGEP as a candidate for providing an answer to the 
well-known question of how it is possible that a simple two-body Schr\"odinger 
equation may represent a solution to a theory as complex as QCD \cite{long, Zalewski}.

A very brief recapitulation of the RGEP is provided here for completeness.
The derivation of $H_\lambda$ begins with a unitary change of the degrees 
of freedom from the bare quarks and gluons in Eq. (\ref{H}) to the effective 
ones. Let $q$ commonly denote the bare operators $b^\dagger$, $d^\dagger$, 
and $a^\dagger$, and their Hermitian conjugates. The operators $q$ are 
transformed by a unitary operator $\cal U_\lambda$ into operators $q_\lambda$ 
that create or annihilate effective particles with identical quantum numbers. 
\begin{equation}
q_\lambda \,\, = \,\, {\cal U}_\lambda \,\, q \,\, {\cal
U}^{\,\dagger}_\lambda \,\, .
\end{equation}
The bare point-like particles in $H$ of Eq. (\ref{H}) correspond to 
$\lambda$ equal infinity. One rewrites the Hamiltonian $H$ in terms of 
$q_\lambda$ and obtains
\begin{equation}
\label{calH} 
H = H_\lambda(q_\lambda) \, .  
\end{equation}
Using ${\cal U}_\lambda$, one has
\begin{equation}
{\cal H}_\lambda \equiv H_\lambda(q) = {\cal U}^{\,\dagger}_\lambda
H {\cal U}_\lambda \, .
\end{equation}
Thus, ${\cal H}_\lambda$ has the same coefficient functions in front
of products of $q$s as the effective $H_\lambda$ has in front of the
unitarily equivalent products of $q_\lambda$s.  Differentiating ${\cal
H}_\lambda$ with respect to $\lambda$, one obtains
\begin{equation}
{\cal H}'_\lambda = - [{\cal T}_\lambda, \, {\cal H}_\lambda] \, ,
\end{equation}
where ${\cal T}_\lambda = {\cal U}^\dagger_\lambda \,{\cal
U}'_\lambda$.  ${\cal T}_\lambda$ is constructed using the notion
of vertex form factors for effective particles.  For example, if an 
operator without a form factor has the structure
\begin{equation}
\hat O_{\lambda} = \int [123] \,\, V_\lambda(1,2,3) \,\,
q^\dagger_{\lambda 1} q^\dagger_{\lambda 2} q_{\lambda 3} \, ,
\end{equation}
then the operator with a form factor is written as $f_\lambda {\hat
O}_{\lambda}$ and has the structure
\begin{equation}
f_\lambda {\hat O}_{\lambda} = \int [123] \,\,
f_\lambda({\cal M}_{12},{\cal M}_3)\,\, V_\lambda(1,2,3) \,\,
q^\dagger_{\lambda 1} q^\dagger_{\lambda 2} q_{\lambda 3} \, .
\end{equation}
Different choices of the function $f_\lambda$ imply different
interactions. The choice adopted in this study is \cite{g_QCD}
\begin{equation} 
\label{f}
f_\lambda({\cal M}_{12},{\cal M}_3) = \exp[-({\cal M}_{12}^2 - {\cal
M}_3^2)^2/\lambda^4)] \, . 
\end{equation}
For any operator $\hat O$ expressible as a linear combination of products 
of creation and annihilation operators, $f \hat O$ contains a form factor 
$f_\lambda({\cal M}_c, {\cal M}_a)$ in front of every product. ${\cal M}_c$ 
and ${\cal M}_a$ stand for the total free invariant masses of the particles 
created (subscript $c$) and annihilated (subscript $a$) by a given product. 

The effective Hamiltonian is defined to have the structure 
\begin{equation}
\label{fG}
H_\lambda = f_\lambda G_\lambda \, ,
\end{equation}
where $G_\lambda$ has to be calculated for given $f_\lambda$. One uses 
${\cal G}_\lambda = G_\lambda(q)$, which is introduced in the same way as 
${\cal H}_\lambda$ in Eq. (\ref{calH}). ${\cal G}_I$ satisfies the 
differential equation
\begin{equation}
\label{G}
{\cal G}_I'\,\, = \,\, \left[ f{\cal G}_I, \, \left\{ (1-f){\cal
G}_I \right\}'_{{\cal G}_0} \right] \,\, , 
\end{equation}
where ${\cal G}_I = {\cal G} - {\cal G}_0$, ${\cal G}_0$ is the part
of $H$ that does not depend on the coupling constant $g$, and the curly 
bracket with subscript ${\cal G}_0$ denotes ${\cal T}$ that solves \cite{RGEP}
\begin{equation}
[{\cal T}, \, {\cal G}_0] = [(1 - f) {\cal G}_I]' \, .
\end{equation}
The initial condition for Eq. (\ref{G}) is that ${\cal G}_\infty = H$,
\begin{equation}
\label{intG}
{\cal G}_\lambda \,\, = \,\, H \,\, + \,\,
\int_\infty^\lambda ds \,\left[ f_s{\cal G}_{Is}, \, \left\{(1-f_s){\cal
G}_{Is} \right\} _{{\cal G}_0} \right] \, .
\end{equation}
This equation shows that one can find the counterterms $X$ in $H$ that remove
regularization dependence from ${\cal G}_\lambda$. ${\cal H}_\lambda = 
f_\lambda {\cal G}_\lambda$ and $H_\lambda$ is obtained by replacing $q$ by 
$q_\lambda$.

${\cal G}_{I \lambda}$ is expanded into a series of terms $\tau_n \sim g^n$,
\begin{equation}
{\cal G}_I \quad = \quad \sum_{n=1}^\infty \, \tau_n \, .
\end{equation}
$\tau_1$ is independent of $\lambda$. Only the term $H_{\psi A \psi}$ needs 
to be discussed here. According to Eq. (\ref{Y}), $\tau_1 = \alpha_{21} + 
\alpha_{12}$, where $\alpha_{21}$ denotes terms that create a gluon and 
$\alpha_{12}$ the terms that annihilate a gluon (the left subscript denotes 
the number of creation and the right subscript the number of annihilation 
operators). The corresponding effective Hamiltonian interaction term is 
obtained by multiplying the integrand in Eq. (\ref{Y}) by $f_\lambda$ and 
transforming $q$s into $q_\lambda$s. 

When one neglects the terms that change the number of particles by more 
than one, $\tau_2=\beta_{11} + \beta_{22}$. Equation (\ref{G}) implies
\begin{equation}
\tau_2' = [\{f'\tau_1\}, f\tau_1] \equiv f_2 \, [\tau_1 \tau_1] \, , 
\end{equation}
with $f_2 = \{f'\}f - f\{f'\}$. The first factor $f$ in $f_2$ refers 
to invariant masses in the first $\tau$ in the square bracket, and 
the second $f$ in $f_2$ is for the second $\tau$. The square bracket 
denotes all connected terms that result from contractions that replace 
products $q_i q_j^\dagger$ by commutators $[q_i, q_j^\dagger]$. The 
solution for $\tau_2$ is then given by
\begin{equation}
\tau_{2 \lambda} = {\cal F}_{2 \lambda} [\tau_1 \tau_1] + \tau_{2
\infty} \, , 
\end{equation}
where $\tau_{2 \infty}$ includes $H_{(\psi\psi)^2}$ and the second-order
mass counterterms from $X$ in Eq. (\ref{H}). ${\cal F}_{2\lambda}$ depends 
on incoming and outgoing momenta in the two vertices generated by the $\tau$s. 
If one labels the three successive configurations of particle momenta by 
$a$, $b$, and $c$, in the sequence $a\tau_{ab} b \tau_{bc} c$, and introduces 
the symbol $uv = {\cal M}^2_{uv} - {\cal M}^2_{vu}$, where ${\cal M}^2_{uv}$ 
denotes the free invariant mass of a set of particles from the configuration 
$u$ that are connected to the particles in the configuration $v$ by the 
interaction $\tau_{uv}$ in the sequence $u \tau_{uv} v$, the vertex form 
factor of Eq. (\ref{f}) in the interaction $\tau_{uv}$ can be written as
\begin{equation}
\label{fab}
f_\lambda({\cal M}_{ab},{\cal M}_{ba}) = \exp{[-(ab^2/\lambda^4)]}
\equiv f_{ab} \, . 
\end{equation}
If one then denotes the parent momentum for the vertex $\tau_{uv}$ by 
$P_{uv}$, and writes $p_{uv}$ in place of $P^+_{uv}$, while all the minus 
components of momenta of the virtual quarks and gluons are given by the 
eigenvalues of ${\cal G}_0 = H_{\psi^2} + H_{A^2}$, 
\begin{equation}
\label{F2}
{\cal F}_2 (a, b, c) = { p_{ba} ba + p_{bc} bc \over ba^2 + bc^2 } \,
\left[ f_{ab} f_{bc} - 1 \right] \, . 
\end{equation}

The second-order perturbation theory renders 
\begin{equation}
\label{Hl}
H_\lambda = T_{q \lambda} + T_{g \lambda} 
          + f_\lambda \left[ Y_{qg \lambda} 
          + V_{q \bar q \lambda} + Z_{q\bar q \lambda} \right] \, .
\end{equation} 
The kinetic energy term for effective quarks is
\begin{equation}
\label{tql}
T_{q \lambda} = 
\sum_{\sigma c} \int [k] {k^{\perp \, 2} + m^2_\lambda \over k^+}
  \left[b^\dagger_{\lambda k \sigma c}b_{\lambda k \sigma c} + 
  d^\dagger_{\lambda k \sigma c }d_{\lambda k \sigma c} \right] \, ,
\end{equation}
where
\begin{eqnarray}
\label{mlambda}
m^2_\lambda & = & m^2_0 + (4/3)\, g^2 \, \int [x\kappa] \,
               \tilde r^2_\delta(x)\,
\sum_{12} |j^\nu_{23} \, \varepsilon^*_{\nu 1}|^2 
\nonumber \\
& \times & \left[{\cal F}_{2 \lambda} (m^2,{\cal M}^2, m^2) 
-{\cal F}_{2 \lambda_0} (m^2,{\cal M}^2, m^2) \right]/k^+_3 \, .
\nonumber \\
& &
\end{eqnarray}
In the order of appearance, $m^2_0$ is the quark mass squared 
that should be present in $T_{q \lambda_0}$ in order to fit data 
for quarkonia, $m^2_0 = m^2 + o(g^2)$. Factor 4/3 comes from color, 
$(N_c^2 - 1)/(2N_c)$. The integration measure is,
\begin{equation}
[x\kappa] = dx \, d^2 \kappa^\perp /[ 16\pi^3 x(1-x)] \,
\end{equation}
where $x = k_1^+/k_3^+$ is the fraction of the quark momentum $k_3$ 
carried by the virtual gluon, and $\kappa^\perp = k_1^\perp - x 
k_3^\perp$ is the relative transverse momentum of the gluon with 
respect to the quark 2. The effective mass does not depend on the 
particle motion. This is a unique property of the RGEP in LF dynamics. 
The small-$x$ regularization factor is 
\begin{equation}
\tilde r_\delta (x) = \, r_\delta (x) \, r_\delta (1-x) \, ,
\end{equation}
where \cite{g_QCD}
\begin{equation}
r_\delta(x) = x^\delta \, \theta(x) \, .
\end{equation}
The middle argument of ${\cal F}_{2 \lambda}$ is 
\begin{equation}
\label{calM}
{\cal M}^2 = (m^2 + \kappa^{\perp \, 2})/(1-x) + \kappa^{\perp \, 2}/x \, ,
\end{equation}
and
\begin{equation}
{\cal F}_{2 \lambda} (m^2,{\cal M}^2, m^2)/k^+_3 =
\left[ f^2_\lambda({\cal M}^2,m^2)-1\right]/({\cal M}^2-m^2) \, .
\end{equation}
The gluon kinetic energy term reads
\begin{equation}
\label{tgl}
T_{g \lambda} = \sum_{\sigma c} \int [k] {k^{\perp \, 2} + \mu^2_\lambda 
\over k^+} \, a^\dagger_{\lambda k \sigma c}a_{\lambda k \sigma c} \, .
\end{equation}
The explicit form of $\mu^2_\lambda$ \cite{RGEP, g_QCD} is not needed here.

The next term in Eq. (\ref{Hl}) is $Y_\lambda = f_\lambda Y_{qg \lambda}$, 
\begin{eqnarray}
\label{yl}
Y_\lambda & = \, g \sum_{123} \, \int[123]  
\,r_\delta(x_{1/3})\,r_\delta(x_{2/3})
\, f_\lambda({\cal M}^2_{12},m^2) \nonumber \\
& \times \left[j_{23} \, b^\dagger_{\lambda 2} 
  a^\dagger_{\lambda 1} b_{\lambda 3} +
  \bar j_{23} \, d^\dagger_{\lambda 2} 
  a^\dagger_{\lambda 1} d_{\lambda 3} + 
h.c. \right] \, .
\end{eqnarray} 

The effective potential term, $V_\lambda = f_\lambda V_{q \bar q 
\lambda}$, originates from the exchange of bare gluons with jumps 
in the invariant mass of intermediate states above $\lambda$.
\begin{equation}
\label{vl}
V_\lambda = -g^2 \,\sum_{1234}\int[1234] 
\, \tilde\delta \, t^a_{12} t^a_{43} V_\lambda(13,24) 
\, b_1^\dagger d_3^\dagger d_4 b_2 \, ,
\end{equation}
where
\begin{eqnarray}
 V_\lambda(13,24) & = & 
 { d_{\mu \nu}(k_5) \over k^+_5} 
 j^\mu_{12} \bar j^\nu_{43}  \,
 f_\lambda({\cal M}^2_{13}, {\cal M}^2_{24})        \nonumber \\
&\times & \left[
  \theta(z)
  \tilde r_\delta (x_{5/1}) \tilde r_\delta (x_{5/4}) \,
  {\cal F}_{2 \lambda} (1, 253, 4) \right.          \nonumber \\
&  +    & \left.
  \theta(-z)
  \tilde r_\delta (x_{5/3}) \tilde r_\delta (x_{5/2})
  \,{\cal F}_{2 \lambda} (3, 154, 2) 
        \right] \, .                                \nonumber \\
&  & 
\end{eqnarray}
\begin{figure}[htb]
\includegraphics[scale=1]{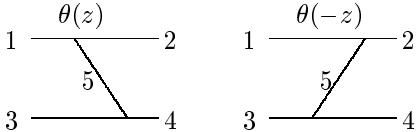}
\caption{\label{fig:oge} Momentum labels in the interaction term 
mediated by the exchange of one high-energy gluon. The same labeling
is used in the exchange of effective low-energy gluon in the next
section and appendices.}
\end{figure}

The sum over polarizations of the intermediate gluon reads
\begin{equation}
\label{dmunu}
d_{\mu \nu}(k_5) = - g_{\mu \nu} + 
{ n^\mu k_5^\nu + k_5^\mu n^\nu \over k_5^+ } \, ,
\end{equation}
where the gluon momentum is 
\begin{equation}
k_5^{+, \perp} = \varepsilon(z) 
\, (k_1^{+, \perp} - k_2^{+, \perp})  \, , 
\end{equation}
and $\varepsilon(z) = \theta(z) - \theta(-z)$,
\begin{equation}
z = (k_1^+ - k_2^+)/(k_1^+ + k_3^+) \, ,
\end{equation}
while $x_5 = |z| = k_5^+/(k_1^+ + k_3^+)$, and 
\begin{equation}
\label{k5shell}
k_5^- = k_5^{\perp \, 2}/ k_5^+ \, .
\end{equation}

The last term in Eq. (\ref{Hl}) is the instantaneous interaction 
between effective quarks, $Z_\lambda = f_\lambda Z_{q\bar q \lambda}$. 
\begin{eqnarray}
\label{zl}
Z_\lambda = -g^2 \sum_{1234}\int[1234] 
\tilde\delta t^a_{12} t^a_{43} Z_\lambda(13,24)
            \, b_{\lambda 1}^\dagger d_{\lambda 3}^\dagger 
               d_{\lambda 4} b_{\lambda 2} , \nonumber \\
\end{eqnarray}
where
\begin{eqnarray}
\label{zzl}
 Z_\lambda(13,24) & = & {1 \over k_5^{+ \, 2}} \, 
j^+_{12} \bar j^+_{34} \,
f_\lambda({\cal M}^2_{13}, {\cal M}^2_{24}) \nonumber \\
& \times & 
[  \theta(z)\tilde r_\delta(x_{5/1}) \tilde r_\delta(x_{5/4})
\nonumber \\
& + & \theta(-z)\tilde r_\delta(x_{5/3}) \tilde r_\delta(x_{5/2})] \, . 
\end{eqnarray}
\begin{figure}[htb]
\includegraphics[scale=1]{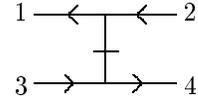}
\caption{\label{fig:sea} Momentum labels in the 
instantaneous gluon interaction term.}
\end{figure}

\section{Eigenvalue equation}
\label{sec:E}
Once $\lambda$ is lowered in perturbation theory to some value 
$\lambda_0$ just above the scale of binding mechanism, the resulting 
$H_{\lambda_0}$ can produce the mass and wave function of a bound 
state of interest in a numerical diagonalization. The basis states 
can be limited to only those that have free invariant masses within 
a range of size about $\lambda$ around the eigenvalue. This has been 
verified numerically in a matrix model with asymptotic freedom and 
bound states \cite{modelafbs, optimization, optimization1}. In that 
calculation, $H_\lambda$ was derived using perturbation theory up to 
6th order. A quite small set of effective basis states with energies 
between 4 MeV and 4 GeV was sufficient to reach accuracy close to 
1\% in the computation of the bound-state energy on the order of 1 
GeV. In great contrast, the initial Hamiltonian of the model coupled 
all states in the entire range between 0.5 KeV and 65 TeV. Preliminary 
estimates performed in Yukawa-like theories also indicate that the 
form factors $f_\lambda$ suppress large momentum changes so strongly 
that the effective dynamics derived in low-order perturbation theory 
receives only small corrections from higher order terms, even when 
the coupling constant is made comparable with 1 \cite{largep}. In 
the case of heavy quarkonia the same strategy should work even more 
accurately than in the Yukawa theory because $\alpha_s$ may be very 
small in comparison to 1. But the weak coupling expansion for 
$H_{\lambda_0}$ produces new interaction terms already in order 
$\alpha_0$. These are derived here. An ansatz for the gluon mass 
allows us then to finesse the structure of the first approximation 
for the resulting $Q\bar Q$-potential. 

The eigenvalue problem for $H_\lambda$ reads
\begin{equation}
\label{evh}
H_\lambda |P\rangle \, = \,  E |P\rangle \, ,
\end{equation}
where $|P\rangle$ denotes an eigenstate of the operators $P_\lambda^+$ 
and $P^\perp_\lambda$ with their eigenvalues denoted by $P^+$ and 
$P^\perp$ (an example of the RGEP construction of the Poincare algebra 
at scale $\lambda$ in quantum field theory is given in \cite{algebra}). 
The eigenvalue $E$ has the form
\begin{equation}
E \, = \, (M^2 + P^{\perp \, 2})/P^+ \, .
\end{equation}
The center-of-mass motion is separated from the binding mechanism,
which is a unique LF-dynamics feature preserved by the RGEP, and 
$P^+$ and $P^\perp$ drop out of the eigenvalue equation. The state 
$|P\rangle$ is written in the effective particle basis as
\begin{equation}
\label{|P>}
|P\rangle = |Q_\lambda \bar Q_\lambda \rangle + 
            |Q_\lambda \bar Q_\lambda g_\lambda \rangle + 
            \, . \, . \, . \, \, .
\end{equation}
The dots denote components with more than three effective particles. Such 
expansion does not apply in the case of the bare particles because those 
interact locally and the interactions disperse probability density to high 
momentum regions and multi-particle sectors \cite{largep}. The wave function 
$\psi_{13}(\kappa^\perp_{13},x_1)$ of the effective valence component
$|Q_\lambda \bar Q_\lambda \rangle$, is introduced by the formula
\begin{eqnarray}
\label{QQ}
|Q_\lambda \bar Q_\lambda \rangle = \int [13] \, P^+\tilde \delta 
\, \psi_{13}(\kappa^\perp_{13},x_1) b^\dagger_{\lambda 1} d^\dagger_{\lambda 3} 
|0\rangle \, ,
\end{eqnarray}
where the quark and anti-quark quantum numbers are labeled with 1 and 3, 
respectively. $\psi_{13}(\kappa^\perp_{13},x_1)$ must have dimension of 
$1/\kappa_{13}^\perp$ for the canonical normalization condition to give 
$\langle P'|P\rangle = P^+\tilde \delta(P-P')$ (quantum numbers of the 
states $|P\rangle$ and $|P'\rangle$ must be the same). The relative 
transverse momentum of two particles, $1$ and $3$, is always defined as 
\begin{equation}
\label{kappa}
\kappa^\perp_{13} = (k_3^+ k_1^\perp - k_1^+ k_3^\perp)/(k_1^+ + k_3^+) \,,
\end{equation}
and $x_1 = k_1^+/P^+ = 1-x_3$. The wave function depends on $\lambda$ 
and quickly vanishes for $|\kappa^{\perp}_{13}| > \lambda$. The 
normalization condition gives,
\begin{equation}
\langle Q_\lambda \bar Q_\lambda |Q_\lambda \bar Q_\lambda \rangle 
= N_{Q\bar Q}(\lambda) \, P^+\tilde \delta(P-P')|_{P'=P} \, ,
\end{equation}
where 
\begin{equation}
N_{Q\bar Q}(\lambda) = \sum_{13}\int[x_1\kappa_{13}]\, 
|\psi_{13}(\kappa^\perp_{13},x_1)|^2 \, .
\end{equation}
The probability of finding other components than the $|Q_\lambda 
\bar Q_\lambda \rangle$ is given by $1 - N_{Q\bar Q}(\lambda)$. 
The value of $N_{Q\bar Q}(\infty)$ is not known but it may be 
close to 0. On the other hand, one expects $N_{Q\bar Q}(\lambda)$ 
to be close to 1 when $m \gg  \Lambda_{QCD}$ and 
\begin{equation}
\label{weakcl}
m \gg \lambda \gg \Lambda_{QCD} \, .
\end{equation}
When the wave function is negligible for relative momenta much 
larger than such $\lambda$, the NR approximation must be accurate in
description of the relative motion of quarks. In addition, when 
$\alpha_\lambda \ll 1$, the Coulomb binding mechanism is expected
to work and the dominant region of momenta should lie around the 
Bohr momentum scale $k_B = \alpha_\lambda \, m$, provided that 
$\lambda \gg k_B$. At the same time, all the fermion spin and 
relativistic correction factors cannot become large (or diverging 
\cite{krp}) in the NR expansion because of the presence of $f_\lambda$ 
\cite{largep}. But the dynamics of the dominant $|Q_\lambda \bar 
Q_\lambda \rangle$ component receives some significant contributions 
from the $|Q_\lambda \bar Q_\lambda g_\lambda \rangle$ component in 
the small-$x_5$ region, since the coupling to the gluon sector grows 
like $\kappa^\perp_5/x_5$ when $x_5 \rightarrow 0$. The gluon component 
may have a negligible contribution to the norm but has to be accounted 
for when $x_5$ is small. 

If one neglected sectors with gluons entirely, the eigenvalue Eq. 
(\ref{evh}) would read
\begin{eqnarray}
\label{evqq}
\left[ T_{q \lambda} + f_\lambda 
\left( V_{q \bar q \lambda} + Z_{q\bar q \lambda}\right) 
\right]|Q_\lambda \bar Q_\lambda \rangle = 
E |Q_\lambda \bar Q_\lambda \rangle \, .
\end{eqnarray}
This equation is mentioned here because an analogous one was considered 
before \cite{gcoh3, QQgcoh, QQ} in a scheme using coupling coherence 
and the absolute lower bound on gluon momenta, $k^+_5 > \delta^+$. 
The equation found in \cite{gcoh3} had a finite limit when $\delta^+ 
\rightarrow 0$. The resulting dynamics contained a logarithmically 
rising potential and reproduced some of the characteristic features of 
the charmonium and bottomonium spectra. This was a considerable success 
in view of how crude were the approximations and the fact that the 
potential derived in \cite{gcoh3} and employed in \cite{QQgcoh, QQ} 
behaved differently in the transverse and longitudinal directions. But 
that strategy could not work in the RGEP approach.

Three reasons can be given now for why the pure $ |Q_\lambda \bar Q_\lambda 
\rangle $ approximation is not allowed in solving the eigenvalue problem 
for $H_\lambda$. Two of them are related to the fact that the coupling 
between the sectors $ |Q_\lambda \bar Q_\lambda \rangle $ and $ |Q_\lambda 
\bar Q_\lambda g_\lambda \rangle $ is proportional to the first power of 
the coupling constant $g$, being mediated by the term $Y_\lambda$ of Eq. 
(\ref{yl}). The first argument is non-perturbative. It is based on the 
result that the matrix models with asymptotic freedom and bound states lead 
to a successful approximation to the eigenvalue equation with $H_\lambda$ 
of second order in $g_\lambda$ if and only if all important matrix elements 
in the properly chosen energy window are accounted for. These certainly 
include matrix elements on the order of $g$ \cite{modelafbs, optimization}. 
The second argument is perturbative, and concerns the evaluation of the 
effective Hamiltonian that acts in the sector $|Q_\lambda \bar Q_\lambda 
\rangle$ alone. When the states $|Q_\lambda \bar Q_\lambda g_\lambda \rangle$ 
are lifted in energy by an amount of order 1, quantum transitions in the 
quark-anti-quark sector that proceed through the states $|Q_\lambda \bar 
Q_\lambda g_\lambda \rangle$ are formally of order $g^2$ and must be 
included when one computes the quark-anti-quark dynamics in a series of 
powers of $g$ up to terms of the explicit order of $\alpha_\lambda$. The 
third argument is based on the fact that Eq. (\ref{evqq}) has a finite
limit when $\delta \rightarrow 0$ only in the leading NR approximation
\cite{RGEP}. Relativistic corrections contain singularities \cite{divx1, 
divx2} and the additional gluon sector has to be taken into account to 
remove them.

The eigenvalue Eq. (\ref{evh}) implies that the $|Q_\lambda \bar 
Q_\lambda g_\lambda \rangle$ component satisfies equation
\begin{eqnarray}
\label{evqqg}
\left[ T_{q \lambda} + T_{g \lambda} + 
       V_{q \bar q g \lambda} - E \right]  
      |Q_\lambda \bar Q_\lambda g_\lambda \rangle =
- Y_\lambda  |Q_\lambda \bar Q_\lambda \rangle \, . \nonumber \\
\end{eqnarray}
$V_{q \bar q g \lambda}$ denotes interactions with sectors with more than 
one gluon and/or additional quark-anti-quark pairs, and the non-abelian 
gluon-quark and quark-anti-quark potentials of order $\alpha_\lambda$. 
The interactions cause a shift in the gluon energy and make the eigenvalue 
equation differ from a similar one for positronium. The idea of seeking 
an ansatz for the shift and building a corresponding first approximation in 
the quark-anti-quark sector may seem completely new, but it patterns QED 
with the exception that there one knows from the outset that the leading 
approximation to a Hydrogen atom or positronium is given by a two-body 
Schr\"odinger equation with a Coulomb potential \cite{CasLep, krp, qed}. 
The NR lattice approach to heavy quarkonia \cite{nrqcd1, nrqcd2} also starts 
from a two-body picture. The key argument is not theoretical but comes from 
the phenomenology of hadrons. Theoretically, an ansatz for the energy shift 
in the sector $|Q_\lambda \bar Q_\lambda g_\lambda \rangle $ is an attempt 
to harness the giant eigenvalue problem for $H$ by turning it into the 
eigenvalue problem for $H_\lambda$ and identifying the corrected Coulomb 
picture that may apply as a first approximation in QCD.

A practical way to increase the invariant mass of the three-body sector 
and preserve the kinematical symmetries of LF dynamics is to add a mass 
$\mu^2$ to $T_{g \lambda}$, using the rules outlined in Section \ref{sec:I}, 
see Eq. (\ref{ansatz1}). Since the divergence in the small-$x$ region 
disappears in the case of positronium when one adds a sector with a 
massless photon, a gluon mass that approaches zero when $x_5 \rightarrow 
0$ can remove the small-$x_5$ divergence in the quarkonium case. The 
rotational symmetry condition on $\mu^2$ can be imposed by demanding that 
the resulting potential in the quark-anti-quark sector is a rotationally 
symmetric function in the center-of-mass variables (see below).

Given a gluon mass ansatz, the whole eigenvalue problem for $H_\lambda$ 
is limited to only two coupled equations,
\begin{eqnarray}
\label{matrix}
(T_q + \tilde T_g)\,|Q\bar Qg\rangle \, + \, Y|Q\bar Q\rangle 
& = & E\,|Q\bar Qg\rangle \, ,\nonumber \\ 
Y\,|Q\bar Qg\rangle \, + \, 
\left[T_q\, + f\left( V_{q \bar q} + Z_{q\bar q}\right)\right]  
|Q\bar Q\rangle & = & E\,|Q\bar Q\rangle  \,  \nonumber \\
& & 
\end{eqnarray}
(the subscript $\lambda$ is omitted). $\tilde T_{g \lambda}$ differs 
from  $T_{g \lambda}$ of Eq. (\ref{tgl}) by replacement of the 
perturbative $\mu^2_\lambda$ by the ansatz $\mu^2$. It is understood 
that $\mu^2$ may depend on $\lambda$. All terms of order $g^2$ in the 
three-body sector are ignored because they do not contribute to the 
dynamics of the $|Q\bar Q\rangle$ component in second-order perturbation 
theory (see below). This dynamics is described by the Hamiltonian 
$H_{Q\bar Q}$ that acts only in the quark-anti-quark sector. One should 
keep in $H_{Q\bar Q}$ terms of formal orders 1, $g$, and $g^2$, when 
the effective Hamiltonian $H_\lambda$ is calculated up to the terms 
of order $g^2$, while $\mu^2 \sim 1$. The bare $g$ is understood to go 
over in higher order calculations to a suitably defined $g_\lambda$ 
\cite{g_QCD, optimization1}. The perturbative expansion is applied only 
in the evaluation of $H_{Q\bar Q}$. Solution for the bound-state spectrum 
of $H_{Q\bar Q}$ is not perturbative.

To evaluate $H_{Q\bar Q}$ as a power series in $g$, one can introduce an 
operator $R$ that expresses the 3-body component through the 2-body one, 
\begin{equation}
|Q\bar Qg\rangle = R|Q\bar Q\rangle \, . 
\end{equation}
Since $Y$ is of order $g$, $R$ is expected to be at least of order $g$. 
If $\hat {\cal P}$ denotes the projector on the $|Q \bar Q \rangle$ sector, 
one has $R = (1-\hat{\cal P})R = R\hat{\cal P}$ and $\hat{\cal P}R = R
(1-\hat{\cal P}) = 0$. The effective Hamiltonian in the $|Q\bar Q\rangle$ 
sector is then given by the formula \cite{Bloch, Wold} (see also \cite{largep} 
concerning the context of RGEP)
\begin{equation}
H_{Q\bar Q} = {1 \over \sqrt{\hat{\cal P} + R^\dagger R}}\,
(\hat{\cal P}+R^\dagger )H_\lambda(\hat{\cal P}+R)\,
{1 \over \sqrt{\hat{\cal P} + R^\dagger R}} \, .
\end{equation}
In the first order in $g$,
\begin{equation}
\label{R}
R T_q - (T_q + \tilde T_g) R = Y \, .
\end{equation}
Consequently, the second-order expression for the matrix 
elements of $H_{Q\bar Q}$ between different states $i$ and $j$ in
the $|Q \bar Q \rangle$ sector, is
\begin{eqnarray}
\label{hqqij}
\langle i|H_{Q\bar Q}|j\rangle = \langle i|
\left[ T_q\, + f\left( V_{q \bar q} + Z_{q\bar q}\right) \right] 
|j\rangle \nonumber \\
+ {1\over 2} \langle i| Y\left( {1\over E_j-T_q -\tilde T_g} +
                    {1\over E_i-T_q -\tilde T_g} \right) Y |j\rangle \, .
\end{eqnarray}
The effective eigenvalue equation for heavy quarkonia, 
$H_{Q\bar Q} |Q \bar Q \rangle = E |Q \bar Q \rangle$,
takes the form
\begin{widetext}
\begin{eqnarray}
\label{schroedinger}
\left[{\kappa^{\perp \, 2}_{13} + m^2_\lambda \over x_1 x_3} + {m_Y^2(1)\over x_1}
                                                  + {m_Y^2(3)\over x_3}  - M^2 \right]
    \psi(\kappa^\perp_{13},x_1) 
  - {4 \over 3}{ g^2 \over 16 \pi^3 } \int {dx_2 d^2 \kappa^\perp_{24} \over x_2 x_4 }
  \, v_\lambda(13,24) \, \psi(\kappa^\perp_{24},x_2) = 0 \, ,
\end{eqnarray}  
where 
\begin{eqnarray}
\label{vschroedinger}
v_\lambda(13,24) & = & \, V_\lambda(13,24) \, + \, Z_\lambda(13,24) \, + \, 
 \, {1 \over 2 x_5 }\,d_{\mu \nu}(k_5) \, j^\mu_{12} \bar j^\nu_{43}  \, w_\lambda(13,24)  \, ,
\end{eqnarray}
and 
\begin{eqnarray}
\label{wlambda}
w_\lambda(13,24) 
& = & 
\left\{ \left[
 { \theta(z)\,
  \tilde r_\delta (x_{5/1}) \tilde r_\delta (x_{5/4}) \,
  f_\lambda (m^2,{\cal M}^2_{52})  
  f_\lambda ({\cal M}^2_{53},m^2)  
\over 
       (\kappa_{13}^{\perp \, 2}+m^2)/x_1 
     - [(\kappa_{13}^\perp-\kappa_{24}^\perp)^2 + \mu^2(2,5,3)]/x_5 
     - (\kappa_{24}^{\perp \, 2}+m^2)/x_2 }
\right. \right.  \nonumber \\
& + &  \left. \left.
 { \theta(-z)\,
  \tilde r_\delta (x_{5/3}) \tilde r_\delta (x_{5/2}) \,
  f_\lambda (m^2,{\cal M}^2_{54})  
  f_\lambda ({\cal M}^2_{51},m^2) 
\over
       (\kappa_{13}^{\perp \, 2}+m^2)/x_3 
     - [(\kappa_{24}^\perp-\kappa_{13}^\perp)^2 + \mu^2(1,5,4)]/x_5 
     - (\kappa_{24}^{\perp \, 2}+m^2)/x_4 }      
\right] \right.  \nonumber \\
& + &  \left. \left[
 { \theta(z)\,
  \tilde r_\delta (x_{5/1}) \tilde r_\delta (x_{5/4}) \,
  f_\lambda (m^2,{\cal M}^2_{52})  
  f_\lambda ({\cal M}^2_{53},m^2)  
\over 
       (\kappa_{24}^{\perp \, 2}+m^2)/x_4 
     - [(\kappa_{13}^\perp-\kappa_{24}^\perp)^2 + \mu^2(2,5,3)]/x_5 
     - (\kappa_{13}^{\perp \, 2}+m^2)/x_3 }   
\right. \right.   \nonumber \\
& + &  \left.\left. 
 { \theta(-z)\,
  \tilde r_\delta (x_{5/3}) \tilde r_\delta (x_{5/2}) \,
  f_\lambda (m^2,{\cal M}^2_{54})  
  f_\lambda ({\cal M}^2_{51},m^2) 
\over
       (\kappa_{24}^{\perp \, 2}+m^2)/x_2 
     - [(\kappa_{24}^\perp-\kappa_{13}^\perp)^2 + \mu^2(1,5,4)]/x_5 
     - (\kappa_{13}^{\perp \, 2}+m^2)/x_1 }                    
\right] \right\} \, . 
\end{eqnarray}
The terms with $\theta(z)$ describe the emission of the gluon by 
the quark and absorption by the anti-quark, while the terms with 
$\theta(-z)$ describe the gluon emission by the anti-quark and 
absorption by the quark, see Fig. \ref{fig:oge}. The first square 
bracket corresponds to the first term in the large round bracket in 
Eq. (\ref{hqqij}), and the second bracket corresponds to the 
second term. The mass terms, $m^2_Y$, originate from the emission 
and re-absorption of the effective gluon by the same quark, in which 
case both terms in the bracket of Eq. (\ref{hqqij}) are equal. The 
mass terms read
\begin{eqnarray}
\label{mY1}
m^2_Y(1) = (4/3)\, g^2 \, \int [x\kappa] \, \tilde r^2_\delta(x)\,
  f^2_\lambda (m^2,{\cal M}^2) \, {j^\nu j^{\mu *} \, 
  d_{\mu \nu}(k)/x_1 \over 
  (\kappa_{13}^{\perp \, 2}+ m^2)/x_1  
- (\kappa_{13}^{\perp \, 2}+{\cal M}_1^2)/x_1 } \, ,   
\end{eqnarray}
where
\begin{eqnarray}
\label{m1}
{\cal M}_1^2 = [\kappa^{\perp \, 2}+ \mu^2(1',5',3)]/x
     + (\kappa^{\perp \, 2}+ m^2)/(1-x) \, , 
\end{eqnarray}
and, 
\begin{eqnarray}
\label{mY3}
m^2_Y(3) = (4/3)\, g^2 \, \int [x\kappa] \, \tilde r^2_\delta(x)\,
  f^2_\lambda (m^2,{\cal M}^2) \, {\bar j^\nu \bar j^{\mu *} \,
  d_{\mu \nu}(k)/x_3 \over 
  (\kappa_{13}^{\perp \, 2}+ m^2)/x_3
- (\kappa_{13}^{\perp \, 2}+{\cal M}_3^2)/x_3 } \, ,   
\end{eqnarray}
where
\begin{eqnarray}
\label{m3}
{\cal M}_3^2 = [\kappa^{\perp \, 2}+ \mu^2(1,5',3')]/x
     + (\kappa^{\perp \, 2}+ m^2)/(1-x) \, .
\end{eqnarray}
\end{widetext}
${\cal M}$ is given by Eq. (\ref{calM}). The subscript $1'$ denotes 
the intermediate quark and $5'$ denotes the intermediate gluon in the 
self-interaction of the effective quark 1, and, similarly, $3'$ and $5'$ 
denote the intermediate anti-quark and gluon in the self-interaction of 
the effective anti-quark 3. 

The gluon four-momentum $k_5$ in the sum over polarizations, i.e., in 
$d_{\mu \nu}(k_5)$ in Eq. (\ref{dmunu}), can be written as
\begin{equation}
k_5^\alpha = \varepsilon(z) q_{ij}^\alpha + n^\alpha \, 
\left[ k^-_5 - \varepsilon(z) q_{ij}^-\right ] /2 \, , 
\end{equation}
where $ij$ refers to quarks 1 and 2, or anti-quarks 3 and 4,
\begin{equation}
q_{ij} = k_i - k_j \, . 
\end{equation}
The quark momentum four-vectors are on the mass shell. Since the 
gluon connects two vertices, one momentum $k_5$ in $d_{\mu\nu}(k_5)$ 
is contracted with the current carried by the quark, and the other 
with the current of the anti-quark. In the self-interactions, both 
momenta are contracted with the same current. The momentum $k_5$ 
contracted with current $j^\alpha_{ij}$ can be expressed through 
$q^\alpha_{ij}$. But the current conservation implies that the terms 
proportional to $q_{ij}$ give zero. Therefore, one can replace Eq. 
(\ref{dmunu}) in the gluon exchange terms by
\begin{eqnarray}
\label{dmn}
& d_{\mu \nu}(k_5) = - g_{\mu \nu} \, + \, n_\mu n_\nu  \nonumber \\   
& \times \, \left[k_5^-  + \varepsilon(z)( k_2^-  -  k_1^-  +  k_3^-  -  k_4^-)/2 \right]
/ k_5^+  \, .
\end{eqnarray}
In the quark self-interaction one has
\begin{equation}
\label{dmnself}
 d_{\mu \nu}(k_{5'}) = - g_{\mu \nu} \, + \, 
 n_\mu n_\nu {k_{5'}^-  + k_{1'}^-  -  k_1^-
 \over k_{5'}^+ }\, ,
\end{equation}
with an analogous result for the anti-quark. 

The terms with the metric $g_{\mu\nu}$ are regular in the small-$x_5$ 
region, while the terms with $n_\mu n_\nu$ are singular. The metric 
terms lead in the well-known way to the Breit-Fermi spin-dependent 
terms with a Coulomb potential. A discussion of the Breit-Fermi 
terms and gluons in the context of QCD can be found in \cite{condensates1} 
and references therein. The singular terms with $n_\mu n_\nu$ are 
independent of the quark spin. It is shown below that the latter 
generate the harmonic force between quarks when combined with the 
fermions' self-interactions, which are also independent of the spin. 
Thus, the harmonic force appears without Breit-Fermi terms. This result 
sets the RGEP approach apart from the models employed in Refs. \cite
{condensates1, pot1, pot2, pot3, pot4, pot5, pot6, pot7}, where one 
had to guess whether a confining potential appeared with or without 
Breit-Fermi terms. The spin-independent harmonic force is akin in 
this respect to the lattice picture and the original charmonium model 
based on the Coulomb force \cite{charm1, charm2, charm3, charm4}. 

In the explicit discussion of singular small-$x$ features of Eq. 
(\ref{schroedinger}) in the next section, all the $g_{\mu \nu}$ terms 
are omitted. The reader should keep their presence in mind until they 
are re-inserted in Section \ref{sec:SE}. The symbols of mass, wave 
function, and potential are provided with a tilde as a reminder about 
the need to include the $g_{\mu \nu}$ terms. Also, expressions for the 
quark masses are simplified by considering from now on only $\lambda 
= \lambda_0$. The subscript 0 indicates that $\lambda = \lambda_0$. 
The ansatz for $\mu^2$ is understood to correspond to $\lambda_0$.
 
With the $g_{\mu \nu}$ terms hidden and $\lambda = \lambda_0$, the 
eigenvalue equation reads
\begin{widetext}
\begin{eqnarray}
\label{ev0}
\left({\kappa^{\perp \, 2}_{13} + m^2_0 \over x_1 x_3} +
{\tilde m^2_1\over x_1} + {\tilde m^2_3\over x_3}  - M^2 \right)
  \tilde \psi(\kappa^\perp_{13},x_1) 
  - {4 \over 3}{ g^2 \over 16 \pi^3 } \int {dx_2 d^2 \kappa^\perp_{24} \over x_2 x_4 }
  \, {j^+_{12} j^+_{43} \over P^{+\, 2} } \, { 1 \over x^2_5 }
  \, \tilde v_0(13,24) \, \tilde \psi(\kappa^\perp_{24},x_2) = 0 \, ,
\end{eqnarray}  
where 
\begin{eqnarray}
\label{v0}
\tilde v_0(13,24) & = &  
f_{13,24} \, [k_5^- + \varepsilon (z)(k_2^- - k_1^- + k_3^- - k_4^-)/2] \,
\left[ \theta(z)  \tilde r_{5/1} \tilde r_{5/4} \, {\cal F}_{1,253,4}   
     + \theta(-z) \tilde r_{5/3} \tilde r_{5/2} \, {\cal F}_{3,154,2} \right] \nonumber \\
& + &  f_{13,24} \, [ \theta(z) \tilde r_{5/1} \tilde r_{5/4}
                    + \theta(-z)\tilde r_{5/3} \tilde r_{5,2}] 
\, + \, {1\over 2 }\,[k_5^- + \varepsilon (z)(k_2^- - k_1^- + k_3^- - k_4^-)/2] 
\, w_0 \, ,
\end{eqnarray}
and the last factor of $w_0 \equiv P^+ w_{\lambda_0}(13,24)$ is abbreviated to
\begin{eqnarray} 
w_0 = 
{ \theta(z) \,
  \tilde r_{5/1} \tilde r_{5/4} \, f_{1,52} f_{53,4}  
  \over  k_1^- - \tilde k^-_5(2,5,3) - k_2^-}
+
{ \theta(-z)\,
  \tilde r_{5/3} \tilde r_{5/2} \, f_{3,54} f_{51,2} 
  \over  k_3^- - \tilde k^-_5(1,5,4) - k_4^-} 
+   
{ \theta(z) \,
  \tilde r_{5/1} \tilde r_{5/4} \, f_{1,52} f_{53,4}  
  \over k^-_4  - \tilde k^-_5(2,5,3) - k_3^- }   
+ 
{ \theta(-z)\,
  \tilde r_{5/3} \tilde r_{5/2} \, f_{3,54} f_{51,2} 
  \over k^-_2  - \tilde k^-_5(1,5,4) - k_1^- } .
\nonumber \\ 
\end{eqnarray}
\end{widetext}
The compact notation includes
\begin{eqnarray}
f_{i,j} \equiv f_{\lambda_0}({\cal M}^2_i,{\cal M}^2_j) \, , \\
\tilde r_{5/i} \equiv \tilde r_\delta (x_{5/i}) \, , \\
{\cal F}_{i, k, j} \equiv {\cal F}_{2 \lambda_0} (i, k, j) \, , \\
\label{k5tilde}
\tilde k^-_5(i,j,k) = [\kappa_5^{\perp \, 2} + \mu^2(i,j,k)]/k^+_5 \, , \\
\kappa^\perp_5 \, = \, \varepsilon(z)\, (\kappa_{13}^\perp - \kappa_{24}^\perp ) \, . 
\end{eqnarray}
The mass terms with the $g_{\mu \nu}$ terms suppressed are 
\begin{eqnarray}
\label{selfi}
\tilde m^2_i = {4\over 3} g^2\int [x\kappa] \, \tilde r^2_\delta(x)\,
f^2_{i,i'5'} {|j^+|^2 \over k_i^{+ \, 2} }\,{1\over x^2}\, 
{ x({\cal M}^2 - m^2) \over m^2  - {\cal M}_i^2 } \, ,  \nonumber \\ 
\end{eqnarray}
for $i=1,3$. ${\cal M}_1$ is given by Eq. (\ref{m1}) and ${\cal M}_3$ 
by Eq. (\ref{m3}). In both cases, ${\cal M}^2$ is given by Eq. 
(\ref{calM}), and the factor $(j^+/k_i^+)^2 = 4(1-x)$.

\section{ Small-$\, x$ behavior }
\label{sec:sx}
All small-$x$ singularities of the eigenvalue Eq. (\ref{schroedinger}) 
are contained in Eqs. (\ref{ev0}), (\ref{v0}), and (\ref{selfi}). We 
first discuss the exchange terms, then the mass terms, and finally the 
net effect of the interplay between these terms. 

The analysis hinges on the properties of the energy of motion of a gluon 
with respect to the parent quark, $p_i^+k^-_5 = x_i\kappa_5^{\perp \, 2}
/x_5$. $p_i$ is the momentum of the parent quark $i$. The momentum $\kappa_5
^\perp$ ranges under the integrals from 0 to $\infty$, while $x_5$ can 
reach 0 (in the mass terms, the integrals are over $\kappa^\perp$ and $x$). 
Appendices \ref{A:Q} and \ref{A:M} provide definitions of all variables 
used in the description of the integrands. The key difficulty is that the 
ratio of two variables of different kinds, $\kappa^\perp$ and $x$, varies 
quickly with a change of any one of them. This complexity is related to 
the power counting rules for the Hamiltonian densities on the LF \cite{long}. 
But the analysis described here concerns only the relative motion of the 
effective particles and it is simplified by taking advantage of the NR limit 
after the finiteness of the small-$x$ dynamics with the gluon mass ansatz is 
established.

The singularity in the effective gluon exchange term is tempered by the 
product of two vertex form factors, $ff$. The form factors vanish 
exponentially fast when $\kappa_5^{\perp \, 2}/x_5 \rightarrow \infty$. 
This prevents $x_5$ from becoming small unless $\kappa_5^\perp$ vanishes 
at least as fast as $\sqrt{x_5}$. Therefore, the measure of integration 
over transverse momenta is on the order of $x_5$ when $x_5 \rightarrow 0$ 
and it reduces the divergence to a logarithmic one. The logarithmic divergence
is taken care of using the gluon mass ansatz. The mechanism of reducing 
singularities to only logarithmic ones does not work in the instantaneous 
interaction term $Z_\lambda (13,24)$ and in the terms in $V_\lambda 
(13,24)$ that come without $ff$ in ${\cal F}_{2\lambda}$. But all the 
terms without $ff$ are independent of $\mu^2$ and the $dx_5/x_5^2$ and 
$dx_5/x_5$ singularities cancel out in them perturbatively \cite{z}. 

In the fermion self-interactions an analogous pattern of the singularities 
occurs. But one has to also consider the size of $m^2_0$. The latter is 
determined by the size of the free ultraviolet-finite part of the quark 
mass counterterm in the initial Hamiltonian of Eq. (\ref{H}). That size is 
related to an ansatz for a gluon mass term in the sectors $|Q g \rangle$ 
and $|\bar Q g\rangle$ in the eigenvalue equations for states with quantum 
numbers of a single fermion, see Appendix \ref{A:M}. Eventually, the gluon 
mass ansatz leads to the result that the single-quark eigenvalue diverges 
logarithmically in the limit $\delta \rightarrow 0$, while the quark 
self-interaction in the quarkonium dynamics becomes finite. The self-interaction 
and effective gluon exchange, both finite due to the chosen behavior of the 
gluon mass ansatz, lead together to the harmonic potential which is described 
in the next section. 

According to the Appendix \ref{A:Q}, the dominant exchange terms in Eq. (\ref{ev0})
read
\begin{eqnarray}
\label{vsimple}
\tilde v_0(13,24) &  =  &  \theta(z)  \, \tilde v_{+\, low} 
                     \, +  \, \theta(-z) \, \tilde v_{-\, low} \, ,
\end{eqnarray}
where $\tilde v_{+\, low}$  is given in Eq. (\ref{v+low}), and
$\tilde v_{-\, low}$  in Eq. (\ref{v-low}). In the limit $x_5 
\rightarrow 0$,
\begin{eqnarray}
\label{limitvplus}
\tilde v_{+\, low} & = & f_{1,52} f_{53,4} \, 
{ \mu^2(2,5,3) \over q^{\perp \, 2} + \mu^2(2,5,3) } \, ,  
\end{eqnarray}
and
\begin{eqnarray}
\label{limitvminus}
\tilde v_{-\, low} & = & f_{3,54} f_{51,2} \,
{ \mu^2(1,5,4) \over q^{\perp \, 2} + \mu^2(1,5,4) } \, . 
\end{eqnarray}
Since $q^{\perp \, 2}$ is on the order of $|z|$, one obtains the result 
that if $\mu^2$ vanishes faster than $q^{\perp \, 2}$, i.e., faster 
than $x_5$, the potential produces a finite effect in the limit of 
$\delta \rightarrow 0$. In the denominators of Eqs. (\ref{limitvplus}) 
and (\ref{limitvminus}) there also appears $q_z^2 = (2mz)^2$, which is 
negligible in comparison to the leading terms on the order of $z$ but 
can be included here on the basis of hindsight to take advantage of the 
NR nature of the quarks' motion with respect to each other. The larger 
is the quark mass $m$ for fixed $\lambda_0$ and the smaller is $\Lambda_
{QCD}$, the more accurate the NR picture actually becomes after the 
small-$x$ divergences are removed. Writing $q_z = q t$, with $q=|\vec q\,|$,
$t = \cos{\theta}$, the singular factor $1/x_5^2$ equals $4m^2/q_z^2 = 
(4m^2/q^2) t^{-2}$. The integration measure $d^3 q$ is proportional to $q^2$ 
and the small-$x$ singularity is actually produced by the angle integration 
$dt/t^2$. $\mu^2$ should vanish for $t \rightarrow 0$ in order to remove 
the singularity. An example of such behavior is used below to provide a 
constructive context for the steps that follow. The final result is not 
sensitive to the details of the example. Given that $\mu^2$ vanishes faster 
than $q^2$, one can write 
\begin{eqnarray}
\label{muq2}
\mu^2(i,5,j) = c^2(i,5,j) \, q^2 \, ,
\end{eqnarray} 
and determine behavior of $c(i,5,j)$ from the condition that 
\begin{eqnarray}
\label{cv}
\tilde c (i,5,j) = { c^2(i,5,j) \over 1 + c^2(i,5,j) }  
\end{eqnarray}
should vanish for $x_5 \rightarrow 0$. 

The only information about the three-particle sector that is available 
in the relativistic construction of $\vec q$ and $c(i,5,j)$ are the 
$\perp$ and $+$ components of the momenta $k_i$, $k_5$, and $k_j$. 
Two physical constraints are used in defining a helpful vector $\vec q$: 
the definition must respect all kinematical LF symmetries (to render 
a boost-invariant description of quarkonia), and it must reduce to 
$q_z = 2mz$ for $z \rightarrow 0$ when $|\vec k_{13}|/m$ and $|\vec 
k_{24}|/m$ approach 0 in the NR limit. A geometrically motivated candidate 
for $\vec q$ is provided by the difference between the square of the 
free invariant mass of three effective particles in the state $|Q\bar 
Q g \rangle$ and the square of the invariant mass of the $Q \bar Q$-pair 
in this state. The difference reads
\begin{eqnarray}
\label{DeltaM}
{\cal M}^2_{i5j} - {\cal M}^2_{ij} = 
{\kappa^{\perp \, 2}_5 + x_5^2 {\cal M}^2_{ij} \over x_5 (1-x_5) } \, .
\end{eqnarray}
Multiplication by $x_5(1-x_5)$ produces an expression that tends in the 
limit of $x_5 \rightarrow 0$ to the three-momentum transfer squared that 
appears in the energy denominators in the small-$x$ dynamics. The components 
of $\vec q$ are therefore defined as $q^\perp = \kappa^\perp_5$ and
\begin{eqnarray}
\label{defqz}
q_z = z {\cal M}_{ij} \, .
\end{eqnarray}
Further analysis of all exchange terms shows that if the ansatz mass $\mu^2$ 
behaves like 
\begin{eqnarray}
\label{dmu}
\mu^2 \sim x_5^{1+\delta_\mu}
\end{eqnarray}
(or like $q^2 x_5^{\delta_\mu}$) with $0 < \delta_\mu < 1/2$, the factors 
$\tilde v_{\pm \, low}$ of Eqs. (\ref{v+low}) and (\ref{v-low}) vanish in 
the limit $x_5 \rightarrow 0$ as $x_5^{\delta_\mu}$ independently of the 
terms in the energy denominators on the order of $x_5^{3/2}$ or smaller.
Thus, the gluon exchange term becomes finite when 
\begin{eqnarray}
\label{ci5j}
c(i,5,j) = \, c(t) \, 
\end{eqnarray}
and $c(t)$ is a function that behaves as
\begin{eqnarray}
\label{ct}
c(t)  =  c  \, |t|^{\delta_\mu/2} \, ,
\end{eqnarray}
for $t \rightarrow 0$, with $c$ a constant. 

With this ansatz, the quark mass terms also become finite in the limit 
$\delta \rightarrow 0$. Appendix \ref{A:M} shows details of how $m_0^2$ 
is chosen in agreement with the physical picture explained in the 
Introduction and at the beginning of this section. Eq. (\ref{selfi}) 
gives $m^2_0 + \tilde m^2_i = m^2 + \delta m^2_i$ with $i=1,3$ and
\begin{eqnarray}
\label{mstildemi}
\delta m_i^2 &=& {4 \alpha_0 \over 3 \pi^2} \, 
\int dx d^2 \kappa^\perp \, \tilde r^2_\delta(x)\, f^2_{\lambda_0}(m^2,{\cal M}^2)\,  
\nonumber \\
& \times &   {{\cal M}^2 - m^2 \over x^2}\, 
\left( {1  \over {\cal M}_0^2 - m^2} - {1  \over {\cal M}_i^2 - m^2} \right)\, .  
\nonumber \\
& & 
\end{eqnarray}
The function ${\cal M}_0^2$ is given by Eq. (\ref{M0phi}) in terms of 
the gluon mass function $\mu_0^2$ that must satisfy the condition 
(\ref{massansatzinequality}). The quark self-energies are positive if
\begin{eqnarray}
\label{positivity}
\mu^2(i,5,j) \,  >  \, \mu_0^2 \, .
\end{eqnarray}
A simple way to satisfy this condition is to set $\mu_0^2=0$. Then,
\begin{eqnarray} 
\label{misimple}
\delta m_i^2  &=& {4 \alpha \over 3 \pi^2} \, 
\int dx d^2 \kappa^\perp \, f^2_{\lambda_0}(m^2,{\cal M}^2) \, {1 \over x^2}\nonumber \\
& \times & 
{ \mu^2(i',5',j) \over \mu^2(i',5',j) + (\kappa^{\perp \, 2}+x^2m^2)/(1-x)} \, ,   \nonumber \\
& &
\end{eqnarray}
where $i=1$ and $j=3$ or vice versa. The factors $r_\delta$ are no longer needed.

The small-$x$ regularization disappears from the quarkonium dynamics
entirely. Finite phenomenological parameters that describe the small-$x$ 
behavior of the gluon mass ansatz, such as $\delta_\mu$ in Eq. (\ref{dmu}), 
become responsible for the regularization of the exchange and self-interaction 
terms, preserving their distinct properties. The mass terms grow when 
$\delta_\mu$ decreases, while the effective gluon exchange potential 
provides a negative contribution that increases in magnitude at similar 
rate and compensates the size of the masses at small momentum transfers. 
The net result is described in the next section.    

\section{ The $Q\bar Q$ Schr\"odinger equation }
\label{sec:SE}
The condition (\ref{weakcl}) validates the NR and weak coupling
limits after the small-$x$ divergences are removed by the gluon 
mass ansatz. This section then identifies the leading structure 
in $H_{Q\bar Q}$ in formal order of $\alpha_0$. The additional 
simplification in the case of small $\alpha_0$ is that the dominant 
interaction in Eq. (\ref{schroedinger}) becomes equal to the well-known 
Coulomb term and one can find the leading correction analytically. 

Equation (\ref{schroedinger}) can be re-written using the relative 
three-momentum variables described in Appendix \ref{A:Q}, see Eq. 
(\ref{kij}). The integration measure is
\begin{equation}
{dx_{24} d^2 k^\perp_{24} \over x_2 x_4 } =
{4 d^3 \vec k_{24} \over {\cal M}_{24}} \, ,
\end{equation}
and Eq. (\ref{schroedinger})  takes the form
\begin{eqnarray}
\label{evsimple}
\left[4(m^2 + \vec k^{\, 2}_{13}) +
{\delta m^2_1\over x_1} + {\delta m^2_3\over x_3} - (2m+B)^2 \right]
\psi(\vec k_{13}) \nonumber \\
 + \int {d^3 k_{24} \over (2\pi)^3\sqrt{m^2+k^2_{24}} } 
\, U(\vec k_{13},\vec k_{24})\, \psi(\vec k_{24}) = 0 \, . \nonumber \\
\end{eqnarray}  
The mass corrections include now the $g_{\mu\nu}$ terms that were
suppressed in the previous section, $\delta m_i^2 = \delta \tilde m_i^2 
+ \delta m^2_g$, and the potential is
\begin{eqnarray}
\label{Ur}
&&U(\vec k_{13},\vec k_{24}) = - {4\over 3} f_{13,24} \, 4\pi \alpha    \nonumber \\
&\times&  \left\{{4\sqrt{x_1x_2x_3x_4} \over x^2_5 } \,  
\left[ \theta(z)  \, \tilde v_{+\, low} 
+  \theta(-z) \, \tilde v_{-\, low} \right] \, + v_g \right\} \, \, , \nonumber \\
\end{eqnarray}  
where $v_g$ denotes the $g_{\mu\nu}$ contribution in the exchange term. 

Since the form factor $f_{13,24}$ cuts off changes of the relative 
momenta above $\lambda_0$ exponentially fast, one can focus on the
eigenstates with lowest $M^2$ and take advantage of the conditions 
$|\vec k_{13}| \ll m$ and $|\vec k_{24}| \ll m$ that are satisfied 
in the entire domain of physically relevant probability distribution. 
For such states, one can expand Eq. (\ref{evsimple}) in powers of 
$\vec k/m$, with the exception of the form factors that are needed 
for convergence. The Coulomb force defines the momentum scale of the 
inverse of the quarkonium Bohr radius, $k_B = r_B^{-1} = \alpha_0 m/2$. 
When $\lambda_0 \gg k_B$, the form factor $f_{13,24}$ does not differ 
from 1 in the dynamically dominant region; $f_{13,24}$ matters only 
when one extends the expansion to high powers of $\vec k/m$. These 
would lead to divergent integrals with Coulomb wave functions and 
require counterterms \cite{CasLep, krp}. The latter are not needed 
here and the lowest terms dominate \cite{largep}. The binding energy, 
$B$, is small in comparison to $m$. Writing the quarkonium mass as 
$M = 2m + B$ and neglecting $\sim B^2/m$, one obtains 
\begin{eqnarray}
\label{evnr}
\left[{\vec k^{\, 2}_{13} \over m} - B
+ {\delta m^2_1\over 2m} + {\delta m^2_3\over 2m} \right]
\psi(\vec k_{13}) \nonumber \\
  + \int {d^3 k_{24} \over (2\pi)^3 } 
     \, V_{Q\bar Q}(\vec k_{13}-\vec k_{24})\, \psi(\vec k_{24}) = 0 \, .
\end{eqnarray}  
The structure of $V_{Q\bar Q}$ and the size of the mass corrections 
$\delta m^2_1$ and $\delta m^2_3$ need explanation. 

The two vertex form factors that appear inside the exchange and mass 
terms in Eq. (\ref{schroedinger}), have arguments given in Appendix 
\ref{A:Q} in Eqs. (\ref{m52}) to (\ref{m51}). When one writes the 
product of the two vertex form factors in the NR limit as $\exp{(-u^2)}$, 
$u$ reads
\begin{eqnarray}
\label{defu}
u = \sqrt{2} \, {m \over \lambda_0} \, {1 \over t} \, {q \over \lambda_0} \, .
\end{eqnarray}
The limit $m/\lambda_0 \gg 1$ enforces $q \ll \lambda_0$, the more so the 
smaller is $t$. The Coulomb binding mechanism is intact for $\lambda_0$  
as small as several times $k_B$ \cite{largep}, which is much smaller than
$m$ in the weak coupling limit. Thus, the momentum transfer $q$ is much 
smaller than $k_B$ in all terms that contain $ff$. These terms become then 
negligible in comparison to the Coulomb term, unless they have singularly 
small denominator factors for small $t$. That is the case for the mass and 
exchange terms when $\delta_\mu$ becomes small. In the presence of the 
$g_{\mu \nu}$ contributions that were omitted in Section \ref{sec:sx}, 
these terms are found as follows.

The Hamiltonian $H_{Q\bar Q}$ has the structure 
\begin{eqnarray}
H_{Q\bar Q} & = & m^2 + \delta m^2(ff,g+n,0) - \delta m^2(ff,g+n,\mu) \nonumber \\
& + & f(1-ff)[(g+n,0)+z] \nonumber \\ 
& + & f(ff)[(g+n,\mu) + z] \, , 
\end{eqnarray}
where $g$ denotes the $g_{\mu \nu}$ terms, $n$ denotes the singular $n_\mu n_\nu$
terms, and $z$ denotes the instantaneous terms. The gluon mass ansatz in
energy denominators is indicated by an extra variable in the brackets, and 0 
says that the gluon mass is 0. The last two terms can be re-arranged as
\begin{eqnarray}
f(1-ff)[(g,0)+(n,0)+z] + \nonumber \\
f(ff)[(g,0) + (g,\mu) - (g,0)+ (n,\mu) + z] \, .
\end{eqnarray}
The contribution of $(n,0)+z$ in the first term vanishes in the leading NR
limit, see Appendix \ref{A:Q}. Two of the terms with $(g,0)$ combine to 
$f(g,0)$, and reduce to the Coulomb term with the Breit-Fermi spin corrections. 
The remaining terms, with $f$ in front also being equivalent to 1,
\begin{eqnarray}
\label{last1}
f(ff)[(g,\mu) - (g,0)+ (n,\mu) + z] \, ,
\end{eqnarray}
add to the Coulomb term and produce together $V_{Q\bar Q}$ in Eq. (\ref{vnr}).
The mass terms can be re-written, in the same fashion, as 
\begin{eqnarray}
\label{last2}
(ff)\delta m^2[(g,0) - (g,\mu)+ (n,0) - (n,\mu)] \, .
\end{eqnarray}
Expressions (\ref{last1}) and (\ref{last2}) show that the exchange
potential and the mass terms have similar structures with opposite 
signs. A change of variables from $x$ and $\kappa^\perp$ to $q_z = x m$ 
and $q^\perp = \kappa^\perp$ in the mass terms produces integrals in 
which the factor $ff$ ensures that $q=|\vec q \, | \ll m$ and one can 
again use the expansion in powers of the ratio of $q/m$. Since the integrands 
are symmetric functions of $q_z$, one can extend the integration to negative 
$q_z$ and divide the result by 2, which produces the same integrands as in 
the exchange terms. Hence,
\begin{eqnarray}
\label{vnr}
V_{Q\bar Q}(\vec q \,) &=& (1 + BF) V_C(\vec q \,) \, + W(\vec q \,) \, ,
\end{eqnarray}
where $BF$ denotes the Breit-Fermi spin-dependent factors,  
\begin{eqnarray}
\label{vc}
V_C(\vec q) &=& - \, {4\over 3} \, { 4\pi \alpha  \over q^2 } \, , \\
W(\vec q \,) & = & \,{4\over 3} \, 4\pi \alpha\,  
\left[{1\over \vec q\,^2} \, - \, {1\over q_z^2}\right] \, 
{\mu^2 \over \mu^2 + \vec q^{\, 2} } \, \nonumber \\
& \times & 
\exp{\left[-2 \left( {m q^2 \over q_z \lambda_0^2} \right)^2\right]} \, ,
\end{eqnarray}  
with $\mu^2 = \theta(z)\,\mu^2(2,5,3) + \theta(-z) \,\mu^2(1,5,4)$, and
\begin{eqnarray}
\label{deltamnr}
{\delta m^2_i \over m} & = & - \, \int {d^3 q \over (2\pi)^3 }\, 
W(\vec q \,) \, ,
\end{eqnarray}
with $\mu^2$ equal $\mu^2(1',5',3)$ for $i=1$ and $\mu^2(1,5',3')$ for $i=3$. 

If the gluon mass ansatz is 0, $W=0$ and the quarkonium dynamics 
reduces to the same as in QED with additional color charge factor 4/3. 
A finite gluon mass ansatz introduces new dynamics which is discussed
in the remaining part of this section.

$W$ is large and negative when $\delta_\mu$ is small. The exchange term 
tends to compensate the positive contribution of the mass terms. This 
can be made transparent by re-writing Eq. (\ref{evnr}) as
\begin{eqnarray}
\label{one}
&&\left[{\vec k^{\, 2} \over m} - B\right] \psi(\vec k) 
+ \int {d^3 q \over (2\pi)^3 } 
     \, (1 + BF) V_C(\vec q \,) \psi(\vec k- \vec q\,)\nonumber \\
&&+ \int {d^3 q \over (2\pi)^3 } 
     \, W(\vec q \,)\, \left[\psi(\vec k - \vec q\,)-\psi(\vec k\,) \right] = 0 \, .
\end{eqnarray}  
There is no need to trace the small relativistic corrections before the 
main NR picture is identified. Only this picture is discussed below.

Since $|\vec q \,|$ in $W$ is constrained to values much smaller than 
$k_B$, one can expand the wave function in the Coulomb region under the 
integral in the Taylor series and consider the lowest terms as candidates
for the first approximation,
\begin{eqnarray}
\psi(\vec k - \vec q) &=& \psi(\vec k)  
- q_i \, {\partial \over \partial k_i} \, \psi(\vec k) \nonumber \\
&+& {1\over 2} \, q_i q_j \, {\partial^2 \over 
\partial k_i \, \partial k_j} \, \psi(\vec k)
+ \,.\,.\,. 
\end{eqnarray}
The terms with odd powers of $\vec q$ average to 0. The bilinear terms 
contain $q^2$ times $(1-t^2)$ times $\cos^2{\cal \phi}$, or $\sin^2{\cal 
\phi}$, for $i = j = 1$, 2, respectively, and $t^2$, for $i= 3$. The 
integral over $\cal \phi$ produces $\pi$ times a vector 
\begin{eqnarray}
\label{vecw}
\vec w(t) = (1-t^2, 1-t^2, 2t^2) \, . 
\end{eqnarray}
The variable $q$ can be changed to $u$ of Eq. (\ref{defu}), and introducing 
the constant
\begin{eqnarray}
\label{defb}
b = {\sqrt{2} m \over \lambda_0^2} \, ,
\end{eqnarray}
one obtains the vector 
\begin{eqnarray}
\label{tau1c}
\vec \tau = \int_0^1 dt \, t (1-t^2) \, \vec w(t) \,\tau (t) \, ,
\end{eqnarray}
\begin{eqnarray}
\label{tau2}
\tau (t) = \int_0^\infty du \,  {(b\mu/t)^2 u^2 \over (b\mu/t)^2 + u^2 } \,
e^{-u^2} \, .
\end{eqnarray}
that appears in the resulting interaction term: 
\begin{eqnarray}
\label{ho}
W_{Q \bar Q} = - \, {4\over 3}\,{ \alpha  \over 2\pi } \,\, b^{-3} \, 
\sum_{i=1}^3 \, \tau_i \, {\partial^2 \over \partial k_i^2 } \, .
\end{eqnarray}
The next non-vanishing terms in the Taylor expansion contain the 
fourth and higher even powers of $\vec q \, $. They are expected 
to be small in the momentum region dominated by the Coulomb dynamics
and do not count around the bottom of the harmonic potential. The
remaining question is if the harmonic approximation can be 
rotationally symmetric.

The interaction $W_{Q \bar Q}$ given by Eq. (\ref{ho}) is rotationally 
symmetric when all components of $\vec \tau$ are equal, or
\begin{eqnarray}
\label{rsymmetry}
\int_0^1 {dt} \, t \,(1-t^2)\, (1-3t^2) \, \tau(t) \, = \, 0 \, .  
\end{eqnarray}  
The function $\tau(t)$ depends on $\mu$ in a limited way because the 
integral over $u$ in Eq. (\ref{tau2}) extends only from 0 to about 
1, $b/t$ is large, and $b\mu/t \gg 1$ produces $\tau(t) = \beta$, 
\begin{eqnarray}
\beta = \sqrt{\pi}/4 \, . 
\end{eqnarray}
Behavior of $\tau(t)$ near $t=0$ does not matter because of the 
factor $t$ in Eq. (\ref{rsymmetry}), and the condition (\ref{dmu}) 
is of little consequence if $\mu^2$ raises quickly from 0 at $t=0$. 
For any ansatz of Eq. (\ref{ct}) with a small $\delta_\mu$,
\begin{eqnarray}
\label{muq}
\tau(t) = {c^2(t) \over 1 + c^2(t)}\,\beta \, ,
\end{eqnarray} 
which is equivalent to the constant $\beta$ if $c(t) \gg 1$ for $t \neq 
0$. The factorization feature is independent of the shape of the RGEP 
form factor $f$ and provides an opportunity to fit $\mu$ analytically 
to satisfy Eq. (\ref{rsymmetry}). Suppose that Eq. (\ref{ct}) is valid 
and 
\begin{eqnarray}
{c^2(t) \over 1 + c^2(t)} \, = c^2 \, t^{\delta_\mu}  \, (1 - \rho t^2) \, .
\end{eqnarray} 
Eq. (\ref{rsymmetry}) is satisfied when 
\begin{eqnarray}
\rho =  \delta_\mu (8+\delta_\mu)/(2+\delta_\mu) \, ,
\end{eqnarray} 
and  
\begin{eqnarray}
c^2(t) = { c^2\,t^{\delta_\mu}  \, (1 - \rho t^2) \over 
1 - c^2\, t^{\delta_\mu}  \, (1 -\rho t^2) } \, ,
\end{eqnarray} 
leads to a rotationally symmetric harmonic oscillator potential.
All components of $\vec \tau$ are equal $\tilde \tau$,
\begin{eqnarray}
\label{tauexample}
\tilde \tau & = & \beta c^2 \left( {1\over 6} \, - \, {\rho \over 12} \right) \, .
\end{eqnarray}
$\rho$ varies between 0 and 17/25 when $\delta_\mu$ varies between
0 and 1/2 in accord with Eq. (\ref{dmu}). In the limit $\delta_\mu 
\rightarrow 0$, $\rho \rightarrow 0$ and $c^2(t) \sim c^2/(1-c^2)$ 
outside the area of very small $t$. In this example, the size of 
the coefficient functions $c(i,5,j)$ depends on the size of the constant 
$c$ and grows to large values when $c \rightarrow 1$. 

Every gluon mass ansatz, $\mu^2$, that is large outside the area of small 
$t$ and vanishes abruptly for $t \rightarrow 0$, leads to a spherically 
symmetric harmonic oscillator potential with the spring constant 
\begin{eqnarray}
k = {4\over 3} \, {\alpha \over \pi} \, b^{-3} \,\tilde \tau \, ,
\end{eqnarray}
and
\begin{eqnarray}
\label{taugeneral}
\tilde \tau  \sim  \beta / 6 \, .
\end{eqnarray}
The number $\beta$ depends on the shape of the RGEP form factor $f$,
which reflects the dependence of the effective dynamics on the RG 
scheme, but $\beta$ is stable for $f_{ab}$s of similar shapes as 
functions of $ab/\lambda^2$, see Eq. (\ref{fab}). 

The first approximation for heavy quarkonium dynamics in position space 
can be defined by the Fourier transform of the eigenvalue equation for 
$H_{Q\bar Q}$, with 
\begin{eqnarray}
\langle \vec r \, | \vec k \, \rangle = \exp{( i \, \vec k \, \vec r \, )} \, .
\end{eqnarray}
This transform exists only in the relative motion variables, since the
motion of the quarkonium as a whole is described in a relativistic
fashion and the relation between the relative motion of quarks and the 
motion of the bound state as a whole does not coincide for large speeds 
with the one known in NR theory. The Schr\"odinger equation reads
\begin{eqnarray}
\label{result}
\left[2m - {\Delta_r \over m} - {4 \alpha\over 3} \left({1\over r} + BF\right) 
 +  {k\over 2} r^2 \right]\psi(\vec r) = M \psi(\vec r) , 
\nonumber \\
\end{eqnarray} 
where $k = m\omega^2/2$, and
\begin{eqnarray} 
\omega &=& \sqrt{ {4\over 3} \, {\alpha \over \pi} } \,\, 
           \lambda \, \left( \lambda \over m \right)^2 \,
           \sqrt{ \beta \over 6 \sqrt{2} }
\end{eqnarray}
The number $[\beta/(6 \sqrt{2})]^{1/2}$ gives $(\pi /1152)^{1/4} \sim 
0.23$, which is large enough for the frequency $\omega$ to fit into 
the ball park of phenomenologically plausible scales when one allows 
sufficiently large $\lambda$ and $\alpha$ with some choice for $m$. 
The observed spectrum of charmonium is known to have an intermediate 
character between the Coulomb and harmonic oscillator spectra. But the 
key problem is to determine the size and direction of corrections that 
need to be included in order to compare the theory with data. The 
inclusion of light quarks requires quantitative understanding of the 
mechanism of chiral symmetry breaking in the effective particle approach 
and a comprehensive phenomenological analysis demands further advance 
in the theory. 

\section{Conclusion}
\label{sec:C}
The Coulomb interaction between quarks in heavy quarkonia is corrected 
by the potential well that is excavated by the one effective gluon 
exchange in the overlapping self-interaction gluon clouds of the quarks. 
At the bottom, the well shape is a quadratic function of the distance 
between the quarks. The resulting harmonic oscillator force plays the 
role of a confining one in a limited range. At distances much larger 
than the Bohr radius the quadratic approximation stops working. Emission 
of additional gluons and pairs of light quarks will further change the 
rate of growth of the potential. The size of these effects should be 
computable in the present approach by evaluating effective Hamiltonians 
order by order in a weak coupling expansion and solving eigenvalue 
problems for them numerically.

The effective particle approach is of interest because it describes the 
relative motion of quarks independently of the speed of the quarkonium 
as a whole. This result is obtained at the price of setting up QCD in 
its Hamiltonian version in LF dynamics, with a host of difficulties in 
the renormalization program that had to be overcome. Further advances in 
the RGEP and methods of solving the eigenvalue equations for Hamiltonians 
$H_\lambda$ are expected to reflect the well known features of interactions 
of relativistic particles. The first approximation for $H_{Q\bar Q}$ can 
be expected to work well in the refined calculations because it appears 
to be largely independent in its structure from the details of the RGEP 
vertex form factors and the gluon mass ansatz. The first approximation 
also appears to involve the least possible degree of complexity as a basis 
around which a meaningful successive approximation scheme can emerge. A 
few percent accuracy in evaluating effective Hamiltonians is known to be 
achievable using essentially the same method in the case of elementary 
matrix models with asymptotic freedom and bound states. 

Since the approach developed here is boost invariant, it can connect 
physical images of hadrons in different frames as soon as the hadron
dynamics is understood in one of them. Although the light quarks are 
expected to behave differently from the heavy ones, one should note 
that the Schr\"odinger equation with $H_\lambda$ does not lead to the 
spread of probability towards large relative momenta and large numbers 
of effective particles. The spread is halted because the interaction 
terms in $H_\lambda$ contain form factors. These form factors are the 
reason for hope that the effective particle expansion may converge.

Aside from QCD, the same scheme for setting up and solving quantum field
theory should be tested in the case of QED. There, the effective mass 
ansatz for virtual photons is much more restricted and small-$x$ effects
are of less significance. On the other hand, QED is not asymptotically 
free and its effective nature requires better understanding. The RGEP 
approach may help in defining QED as an effective theory. But one needs 
to first verify if perturbation theory with $H_\lambda$ can produce 
covariant $S$-matrix in QED in orders higher than second.

\appendix
\section{ Regularized LF Hamiltonian of QCD }
\label{A:H}
The canonical LF Hamiltonians of gauge theories, similar to the 
Hamiltonians in the infinite momentum frame \cite{KS, BKS, S}, are
well known \cite{Yan, Banff}, and extensive literature exists on 
the light-like axial gauges \cite{Bassetto, SriBro}. The Hamiltonian 
given below is further specified by inclusion of the ultraviolet and 
small-$x$ regularization factors that render a computable operator. 
This means that $H$ does not require a separate regularization 
prescription for evaluating loop integrals. The same regularized 
Hamiltonian was used in \cite{g_QCD} but the quark terms needed here 
were not explicitly given there. $H$ is supplied with counterterms 
$H_{\Delta\delta}$. Their structure is known from considerations 
similar to \cite{long}. Details can be calculated using RGEP. The 
initial Lagrangian is 
\begin{equation}
{\cal L} = \bar \psi(i\hspace{-4pt}\not\!\!D - m)\psi - 
{1\over 4}F^{\mu\nu a}F_{\mu\nu}^a \, ,
\end{equation}
with one flavor of quarks of mass $m$, 
\begin{eqnarray}
F^{\mu\nu} = -i[D^\mu,D^\nu]/g \, , 
\end{eqnarray}
and 
\begin{eqnarray}
D_\mu = \partial_\mu + ig A_\mu \, , 
\end{eqnarray}
where $ A = A^a t^a$, $[t^a,t^b] = i f^{abc} t^c $, and $Tr(t^a t^b) = 
\delta^{ab}/2$. The classical Nether generator of evolution in $x^+$ 
takes the form (the Gauss law constraint is formally solved in $A^+=0$ 
gauge and the counterterms are added as the last term from hindsight),
\begin{eqnarray}
H &=& H_{\psi^2} + H_{A^2} + H_{A^3} + H_{A^4} + H_{\psi A \psi} \nonumber \\
&+& H_{\psi A A \psi} + H_{[\partial A A]^2} + H_{[\partial A A](\psi\psi)} 
+ H_{(\psi\psi)^2} \nonumber \\
&+& H_{\Delta \delta} \, ,
\end{eqnarray}
where each term is an integral over the LF hyperplane,
\begin{eqnarray} 
H_i = \int dx^- d^2 x^\perp {\cal H}_i  \, ,
\end{eqnarray}
and,
\begin{eqnarray} 
{\cal H}_{\psi^2} 
&=& {1\over 2} \bar \psi \gamma^+ {-\partial^{\perp \, 2} + 
m^2 \over i\partial^+} \psi \, , \\
{\cal H}_{A^2} &=& 
- {1\over 2} A^\perp (\partial^\perp)^2 A^\perp \, , \\
{\cal H}_{A^3} &=& 
g \, i\partial_\alpha A_\beta^a [A^\alpha,A^\beta]^a \, , \\
{\cal H}_{A^4} &=& 
- {1\over 4} g^2 \, [A_\alpha,A_\beta]^a[A^\alpha,A^\beta]^a \, , \\
{\cal H}_{\psi A \psi} &=& 
g \, \bar \psi \hspace{-4pt}\not\!\!A \psi \, , \\
{\cal H}_{\psi A A \psi} &=& 
{1\over 2}g^2 \, \bar \psi \hspace{-4pt}\not\!\!A 
{\gamma^+ \over i\partial^+} \hspace{-4pt}\not\!\!A \psi \, , \\ 
{\cal H}_{[\partial A A]^2} &=& {1\over 2}g^2 \, 
  [i\partial^+A^\perp,A^\perp]^a {1 \over (i\partial^+)^2 }
  [i\partial^+A^\perp,A^\perp]^a \, , \nonumber \\
& & \\
{\cal H}_{[\partial A A](\psi\psi)} &=& 
g^2 \, \bar \psi \gamma^+ t^a \psi {1 \over (i\partial^+)^2 }
[i\partial^+A^\perp,A^\perp]^a \, , \\
{\cal H}_{(\psi\psi)^2} &=& 
{1\over 2}g^2 \,
\bar \psi \gamma^+ t^a \psi {1 \over (i\partial^+)^2 }
\bar \psi \gamma^+ t^a \psi \, .
\end{eqnarray}
A quantum Hamiltonian is introduced by expanding the fields 
into Fourier components at $x^+=0$ and imposing commutation 
relations on the latter. They define creation and annihilation 
operators for bare particles. 
\begin{eqnarray}
 \psi = \sum_{\sigma c} \int [k] 
 \left[ \chi_c u_{k\sigma} b_{k\sigma c} e^{-ikx} + 
        \chi_c v_{k\sigma} d^\dagger_{k\sigma c} e^{ikx}
 \right] \, . \nonumber \\
\end{eqnarray}
The integration measure is 
\begin{eqnarray}
[k] = { \theta(k^+)\, k^+ \, d^2 \, k^\perp \over 16\pi^3 k^+ } \, .
\end{eqnarray}
\begin{eqnarray}
\{b_{k\sigma c},b^\dagger_{k'\sigma' c'}\} 
&=&\{d_{k\sigma c},d^\dagger_{k'\sigma' c'}\} \nonumber \\
&=& 16\pi^3 k^+ \delta_{\sigma_\sigma'} \delta_{cc'}
\delta^3(k-k') \, .
\end{eqnarray}
$ \delta^3(k-k') = \delta(k^+-k'^+) \delta(k^1-k'^1) 
\delta(k^2-k'^2) $. The creation and annihilation 
operators have the power-counting dimension $1/k^\perp$
(the same result holds for gluons, see below). The spinors
are given by $u_{k\sigma} = B(k,m) u_{\sigma}$, and 
$v_{k\sigma} = B(k,m) v_{\sigma}$, where $v_{\sigma} = 
C u^*_{\sigma} = i\gamma^2 u^*_{\sigma} $. $u_{\sigma}$ and 
$v_{\sigma}$ are the spinors for the fermions at rest 
in the arbitrarily chosen frame of reference where the 
quantization procedure is introduced. The matrix $B(k,m)$
represents the LF boost that turns a particle with mass $m$ 
at rest to a moving one that has the momentum $k$, $k^2 = m^2$,
\begin{eqnarray}
B(k,m) = {1\over \sqrt{k^+ m}} \, 
  [ \Lambda_+ k^+ + \Lambda_- (m + k^\perp \alpha^\perp)] \, ,
\end{eqnarray}
where $\Lambda_\pm = \gamma_0 \gamma^\pm/2$. This matrix mixes
$k^+$ with $m$ and $k^\perp$. But the second term, of the type 
$\sqrt{k^\perp/k^+}$ when one counts $m$ and $k^\perp$ as similar,
results only from writing the interaction terms in a way that is 
short and convenient in calculations. The independent degrees of 
freedom, $\psi_+ = \Lambda_+ \psi$, contain only the parts proportional 
to $\sqrt{k^+/m}$. Thus, $\psi_+$ has the dimension of $k^{\perp \, 2} 
\sqrt{k^+} b$, which in the position space on the LF leads to 
$(x^\perp \sqrt{x^-})^{-1}$ \cite{long} if $b \sim 1/k^\perp$. 
The spinors at rest are,
\begin{eqnarray}
u_{\sigma} = \sqrt{2m}\left[ \begin{array}{c} \chi_\sigma \\ 
                              0 \end{array}\right] \, ,  \\
v_{\sigma} = \sqrt{2m}\left[ \begin{array}{c} 0 \\ 
                  \xi_{-\sigma} \end{array}\right] \, ,
\end{eqnarray}
where $\xi_{-\sigma}=-i\sigma_2\chi_\sigma = \sigma\chi_{-\sigma}$.
The gluon field at $x^+ =0$ is
\begin{eqnarray}
A^\mu = \sum_{\sigma c} \int [k] \left[ t^c \varepsilon^\mu_{k\sigma} 
  a_{k\sigma c} e^{-ikx} + t^c \varepsilon^{\mu *}_{k\sigma} 
  a^\dagger_{k\sigma c} e^{ikx}\right] \, , \nonumber \\
\end{eqnarray}
and the commutation relations read,
\begin{eqnarray}
\left[ a_{k\sigma c}, a^\dagger_{k'\sigma' c'}\right] 
= 16\pi^3 k^+ \delta_{\sigma_\sigma'} \delta_{cc'}
  \delta^3(k-k') \, .
\end{eqnarray} 
$a$ and $a^\dagger$ have the dimension $1/k^\perp$. The polarization 
four-vectors are introduced using LF boosts as for fermions, 
\begin{eqnarray}
\varepsilon^\mu_{k\sigma} = (\varepsilon^+_{k\sigma}=0,
\varepsilon^-_{k\sigma}= 2k^\perp \varepsilon^\perp_\sigma/k^+,
\varepsilon^\perp_\sigma) \, ,
\end{eqnarray}
except that the boosts are applied to the polarization vectors 
$\varepsilon^\mu_\sigma = (0, 0, \varepsilon^\perp_\sigma)$ that 
correspond to the selected state of a gluon moving along the 
$z$-axis \cite{Wigner}. Terms that contain the ratio $k^\perp/k^+$, 
which mix the transverse and longitudinal momenta, are again only a 
shorthand notation for writing interactions. The independent transverse 
field $A^\perp$ contains only polarization vectors $\varepsilon
^\perp_\sigma $ that have dimension 1. $A^\perp$ has dimension of 
$k^{\perp \, 2} a$, which matches the required $1/x^\perp$ on the 
LF \cite{long} when $a \sim 1/k^\perp$, as promised.

The kinetic energy operator for quarks, $H_{\psi^2}$, is given
in Eq. (\ref{tquark}), and for gluons, $H_{A^2}$, is given in
Eq. (\ref{tgluon}). The triple-gluon interaction reads
\begin{eqnarray}
H_{A^3} & = & \sum_{123}\int[123]
  \tilde\delta(p^\dagger - p)\,\tilde r_{\Delta\delta}(3,1) \nonumber \\
& \times &  \left[g\,Y_{123}\, a^\dagger_1 a^\dagger_2 a_3 + 
  g\,Y_{123}^*\, a^\dagger_3 a_2 a_1 \right] \, .
\end{eqnarray}
The symbols introduced in this operator occur in all other terms
and require explanation for completeness, see also \cite{g_QCD}. 
The conservation of momentum in the interaction vertices is enforced 
by the factor
\begin{eqnarray}
\tilde\delta(p^\dagger - p) = 16 \pi^3 \delta^3
\left[\sum_{a^\dagger}p_{a^\dagger}-\sum_{a}p_{a}\right] \, .
\end{eqnarray}
The regularization factors are given by
\begin{eqnarray}
\tilde r_{\Delta\delta}(p,d) & = & r_{\Delta\delta}(p,d)
\,r_{\Delta\delta}(p,p-d) \, ,
\end{eqnarray}
where
\begin{eqnarray}
r_{\Delta\delta}(p,d) & = & r_\Delta(\kappa^{\perp \, 2}_{d/p})
  \,r_\delta(x_{d/p})\,\theta(x_{d/p}) \, .
\end{eqnarray}
The symbol $p$ refers to the parent momentum (half of the sum of 
all momenta of all particles coupled in a vertex), and $d$ to the
daughter particle momentum, i.e., the momentum of the particle 
emitted or absorbed in the vertex. The arguments of the regularization 
factors are defined by
\begin{eqnarray}
  \kappa^\perp_{d/p} & = & k^\perp_d - x_{d/p} k^\perp_p \, , \\
  x_{d/p} & = & k^+_d/k^+_p \equiv x_d/x_p \, .
\end{eqnarray}
The functions used here are \cite{g_QCD} 
\begin{eqnarray}
r_\Delta(\kappa^{\perp \, 2}) & = & \exp{[-\kappa^{\perp \, 2}/\Delta^2]} \, , \\
r_\delta(x)  & = & \theta(x - \epsilon) x^\delta \, , 
\end{eqnarray}
and $\epsilon/\delta$ tends to 0 when $\delta \rightarrow 0$.
The gluon spin vertex factor reads
\begin{eqnarray}
  Y_{123} & = & i f^{c_1 c_2 c_3}  \\
& \times & \left[ 
  \varepsilon_1^*\varepsilon_2^* \cdot \varepsilon_3\kappa 
  - 
  \varepsilon_1^*\varepsilon_3 \cdot \varepsilon_2^*\kappa {1\over x_{2/3}} 
  - 
  \varepsilon_2^*\varepsilon_3 \cdot \varepsilon_1^*\kappa {1\over x_{1/3}} 
  \right] \, . \nonumber 
\end{eqnarray}
The simplified notation means: $\varepsilon \equiv \varepsilon^\perp$,
$\kappa \equiv \kappa^\perp_{1/3}$. The quartic gluon vertex is
\begin{widetext}
\begin{eqnarray}
  H_{A^4} = \sum_{1234}\int[1234]
  \tilde\delta(p^\dagger - p)\,{g^2 \over 4}\, \left[ 
  \Xi_{A^4 \, 1234} a^\dagger_1 a^\dagger_2 a^\dagger_3 a_4 + 
  X_{A^4\,1234} a^\dagger_1 a^\dagger_2 a_3 a_4 +
  \Xi^*_{A^4\,1234} a^\dagger_4 a_3 a_2 a_1 \right] \, ,
\end{eqnarray}
where 
\begin{eqnarray}
 \Xi_{A^4 \, 1234} & = & {2\over 3}[  \tilde r_{1+2,1} \tilde r_{4,3}\,
  (\varepsilon_1^*\varepsilon_3^* \cdot \varepsilon_2^*\varepsilon_4 - 
  \varepsilon_1^*\varepsilon_4 \cdot \varepsilon_2^*\varepsilon_3^*)\,
  f^{a c_1 c_2}f^{a c_3 c_4}  \nonumber \\
& + & \tilde r_{1+3,1} \tilde r_{4,2}\,
  (\varepsilon_1^*\varepsilon_2^* \cdot \varepsilon_3^*\varepsilon_4 - 
  \varepsilon_1^*\varepsilon_4 \cdot \varepsilon_2^*\varepsilon_3^*)\,
  f^{a c_1 c_3}f^{a c_2 c_4} \nonumber \\
& + &  \tilde r_{3+2,3} \tilde r_{4,1}\,
  (\varepsilon_1^*\varepsilon_3^* \cdot \varepsilon_2^*\varepsilon_4 - 
  \varepsilon_3^*\varepsilon_4 \cdot \varepsilon_2^*\varepsilon_1^*)\,
  f^{a c_3 c_2}f^{a c_1 c_4}] \, , 
\end{eqnarray}
\begin{eqnarray}
 X_{A^4 \, 1234} & = & \tilde r_{1+2,1} \tilde r_{3+4,3}\,
  (\varepsilon_1^*\varepsilon_3 \cdot \varepsilon_2^*\varepsilon_4 - 
  \varepsilon_1^*\varepsilon_4 \cdot \varepsilon_2^*\varepsilon_3)\,
  f^{a c_1 c_2}f^{a c_3 c_4} \nonumber \\ 
& + & [\tilde r_{3,1} \tilde r_{2,4}+\tilde r_{1,3} \tilde r_{4,2}] 
  (\varepsilon_1^*\varepsilon_2^* \cdot \varepsilon_3 \varepsilon_4 - 
  \varepsilon_1^*\varepsilon_4 \cdot \varepsilon_2^*\varepsilon_3)\,
  f^{a c_1 c_3}f^{a c_2 c_4} \nonumber \\
& + & [\tilde r_{3,2} \tilde r_{1,4} + \tilde r_{2,3} \tilde r_{4,1}] \,
  (\varepsilon_1^*\varepsilon_2^* \cdot \varepsilon_3^*\varepsilon_4 - 
  \varepsilon_1^*\varepsilon_3 \cdot \varepsilon_2^*\varepsilon_4)\,
  f^{a c_1 c_4}f^{a c_2 c_3} \, .
\end{eqnarray}
The abbreviated notation for the regularization factors is $\tilde r_{p,d} 
\equiv \tilde r_{\Delta\delta}(p,d)$. Denoting $t^a_{ij} \equiv \chi^\dagger_{ic} 
t^a \chi_{jc}$, the quark-gluon coupling is given by
\begin{eqnarray}
 H_{\psi A \psi} = \sum_{123}\int[123]
  \tilde\delta(p^\dagger - p)\,\tilde r_{3,1}\,g\, 
\left[
  \bar u_2 \not\!\varepsilon_1^* u_3 \, t^1_{23} \, b^\dagger_2 a^\dagger_1 b_3 
- \bar v_3 \not\!\varepsilon_1^* v_2 \, t^1_{32} \, d^\dagger_2 a^\dagger_1 d_3 
+ \bar u_1 \not\!\varepsilon_3   v_2 \, t^3_{12} \, b^\dagger_1 d^\dagger_2 a_3 
+ h.c. \right], 
\end{eqnarray}
where the spin vertex factors are
\begin{eqnarray}
\bar u_2 \not\!\varepsilon_1^* u_3 & = & \sqrt{x_3/x_2}\,\, 
  \chi_2^\dagger\,
  \left[ i(\kappa^\perp_{1/3} \times \varepsilon_1^{*\perp})^3
  \sigma^3 + {x_2 + x_3 \over x_1} \,
  \kappa^\perp_{1/3}\varepsilon_1^{*\perp}
  - m_3 {x_1\over x_3}\,
  \sigma^\perp\varepsilon_1^{*\perp}\,\sigma^3 \right]\chi_3 \, , \\
\bar v_3 \not\!\varepsilon_1^* v_2 & = & \sqrt{x_3/x_2}\,\,
  \xi_{-3}^\dagger\,
  \left[-i(\kappa^\perp_{1/3} \times \varepsilon_1^{*\perp})^3
  \sigma^3 + {x_2 + x_3 \over x_1} \,
  \kappa^\perp_{1/3}\varepsilon_1^{*\perp}
  - m_3 {x_1\over x_3}\,
  \sigma^3 \sigma^\perp\varepsilon_1^{*\perp} \right]\xi_{-2} \, , \\
\bar u_1 \not\!\varepsilon_3 v_2  & = & \sqrt{ x_3/x_1}\sqrt{x_3/x_2}\,\, 
  \chi_1^\dagger \,
  \left[-i(\kappa^\perp_{1/3} \times \varepsilon_3^\perp)^3
  + {x_1 - x_2 \over x_3} \,
  \kappa^\perp_{1/3}\varepsilon_3^\perp\,\sigma^3
  - m_1 \sigma^\perp\varepsilon_3^\perp \right]\xi_{-2} \, .
\end{eqnarray}
The instantaneous fermion interaction reads 
\begin{eqnarray}
  H_{\psi A A \psi} = \sum_{1234}\int[1234]
  \tilde\delta(p^\dagger - p)\,(g^2/2)\,2\sqrt{x_1 x_4} 
  \,\, \cdot \,\, \{\} \, ,
\end{eqnarray}
where the curly brackets $\{\}$ contain the operators ordered 
according to the rule $b^\dagger d^\dagger a^\dagger a d b$;  
\begin{eqnarray}
\{\} \, & = & \, \tilde r_{1+2,1} \tilde r_{3+4,3}\,
  \left[
  t^{23}_{14}\, 
  {\chi_1^\dagger[2^* 3]\chi_4 \over x_1 + x_2}\,
  b_1^\dagger a_2^\dagger a_3 b_4 
  +
  t^{23}_{14}\, 
  {\xi_{-1}^\dagger[2 3^*]\xi_{-4} \over x_1 + x_2}
  d_4^\dagger a_3^\dagger a_2 d_1  
  \right]\, \nonumber \\
& + &
 [\tilde r_{1,2} \tilde r_{4,3} + \tilde r_{2,1} \tilde r_{3,4}]\,
  \left[
  t^{23}_{14}\, 
  {\chi_1^\dagger[2 3^*]\chi_4 \over x_1 - x_2}\,
  b_1^\dagger a_3^\dagger a_2 b_4 
  +
  t^{23}_{14}\, 
  {\xi_{-1}^\dagger[2^* 3]\xi_{-4} \over x_1 - x_2}
  d_4^\dagger a_2^\dagger a_3 d_1
  \right]\, \nonumber \\ 
& + &
 \left[\tilde r_{3,4} \tilde r_{1+2,1}\,
  t^{23}_{14}\, 
  {\chi_1^\dagger[2^* 3]\sigma^3\xi_{-4} \over x_1 + x_2}\,
  b_1^\dagger d_4^\dagger a_2^\dagger a_3  + h.c. \right] 
  +
  \left[\tilde r_{2,1} \tilde r_{3+4,3}\,
  t^{23}_{14}\, 
  {\chi_1^\dagger[2 3^*]\sigma*3\xi_{-4} \over x_1 - x_2}\,
  b_1^\dagger d_4^\dagger a_3^\dagger a_2  + h.c \right] \nonumber \\
& + &
  \left[{1\over 2}\left(
  [\tilde r_{1,2} \tilde r_{3,4} + \tilde r_{2,1} \tilde r_{4,3}]\,
  t^{23}_{14}\, 
  {\chi_1^\dagger[2 3]\sigma^3\xi_{-4} \over x_1 - x_2} +
  [\tilde r_{1,3} \tilde r_{2,4} + \tilde r_{3,1} \tilde r_{4,2}]\,
  t^{32}_{14}\, 
  {\chi_1^\dagger[3 2]\sigma^3\xi_{-4} \over x_1 - x_3}
  \right)
  b_1^\dagger d_4^\dagger a_2 a_3 + h.c.\right] \nonumber \\
& + & 
  \left[
  {1\over 2}
  \left(
  \tilde r_{4,3} \tilde r_{1+2,1}\,
  t^{23}_{14}\, 
  {\chi_1^\dagger[2^* 3^*]\chi_4 \over x_1 + x_2}\,
  +
  \tilde r_{4,2} \tilde r_{1+3,1}\,
  t^{32}_{14}\, 
  {\chi_1^\dagger[3^* 2^*]\chi_4 \over x_1 + x_3}\,
  \right)
  b_1^\dagger a_2^\dagger a_3^\dagger b_4  + h.c. \right] \nonumber \\
& + &
  \left[
  {1\over 2}
  \left(
  \tilde r_{1,2} \tilde r_{3+4,4}\,
  t^{23}_{14}\, 
  {\xi_{-1}^\dagger[2^* 3^*]\xi_{-4} \over x_1 - x_2}\,
  +
  \tilde r_{1,3} \tilde r_{2+4,4}\,
  t^{32}_{14}\, 
  {\xi_{-1}^\dagger[3^* 2^*]\xi_{-4} \over x_1 - x_3}
  \right)
  d_4^\dagger a_2^\dagger a_3^\dagger d_1 + h.c. \right]
\end{eqnarray}
The symbols mean: $t^{ab}_{ij} = \chi_{ic}^\dagger 
t^a t^b \chi_{jc}$, and $[\alpha\beta] = \varepsilon_\alpha^\perp 
\varepsilon_\beta^\perp + i (\varepsilon_\alpha^\perp 
\times \varepsilon_\beta^\perp)^3 \sigma^3$. The star,
$*$, means that the corresponding polarization vector is
complex conjugated: $*  \rightarrow \varepsilon^* $.
The quartic gluon term with derivative reads
\begin{eqnarray}
H_{[\partial A A]^2} = \sum_{1234}\int[1234]
  \tilde\delta(p^\dagger - p)\,g^2\, \left[ 
  \left(
  \Xi_{[\partial A A]^2 \, 1234} a^\dagger_1 a^\dagger_2 a^\dagger_3 a_4 
  + h.c. \right) +
  X_{[\partial A A]^2\,1234} a^\dagger_1 a^\dagger_2 a_3 a_4 \right] \, , 
\end{eqnarray}
where
\begin{eqnarray}
 \Xi_{[\partial A A]^2\, 1234} & = & - {1\over 6}\left[ \tilde r_{1+2,1} \tilde r_{4,3}\,
  \varepsilon_1^*\varepsilon_2^* \cdot \varepsilon_3^*\varepsilon_4\, 
  {(x_1 - x_2)(x_3+x_4)\over (x_1 + x_2)^2}\, 
  f^{a c_1 c_2}f^{a c_3 c_4} \right.\nonumber \\
& + & \left.
  \tilde r_{1+3,1} \tilde r_{4,2}\,
  \varepsilon_1^*\varepsilon_3^* \cdot \varepsilon_2^*\varepsilon_4\, 
  {(x_1 - x_3)(x_2+x_4)\over (x_1 + x_3)^2}\,
  f^{a c_1 c_3}f^{a c_2 c_4} \right. \nonumber \\
& + & \left.
  \tilde r_{3+2,3} \tilde r_{4,1}\,
  \varepsilon_3^*\varepsilon_2^* \cdot \varepsilon_1^*\varepsilon_4\, 
  {(x_3 - x_2)(x_1+x_4)\over (x_3 + x_2)^2}\,
  f^{a c_3 c_2}f^{a c_1 c_4} \right]  \, , 
\end{eqnarray}
\begin{eqnarray}
 X_{[\partial A A]^2\, 1234} & = & {1\over 4} \left[ \tilde r_{1+2,1} \tilde r_{3+4,3}\,
  \varepsilon_1^*\varepsilon_2^* \cdot \varepsilon_3 \varepsilon_4\, 
  {(x_1 - x_2)(x_3 - x_4)\over (x_1 + x_2)^2}\, 
  f^{a c_1 c_2}f^{a c_3 c_4} \right. \nonumber \\
& - & \left.
  [\tilde r_{3,1} \tilde r_{2,4}+\tilde r_{1,3} \tilde r_{4,2}]\,
  \varepsilon_1^*\varepsilon_3 \cdot \varepsilon_2^* \varepsilon_4\, 
  {(x_1 + x_3)(x_2 + x_4)\over (x_2 - x_4)^2}\,
  f^{a c_1 c_3}f^{a c_2 c_4} \right.  \nonumber \\
& - & \left.
  [\tilde r_{3,2} \tilde r_{1,4} + \tilde r_{2,3} \tilde r_{4,1}] \,
  \varepsilon_1^*\varepsilon_4 \cdot \varepsilon_2^*\varepsilon_4\, 
  {(x_2 + x_3)(x_1+x_4)\over (x_1 - x_4)^2}\,
  f^{a c_1 c_4}f^{a c_2 c_3} \right] \, .
\end{eqnarray}
The instantaneous gluon interaction between quarks and gluons reads
\begin{eqnarray}
H_{[\partial A A](\psi\psi)} & = & \sum_{1234}\int[1234] 
  \tilde\delta(p^\dagger - p)\ g^2\,if^{a12}\,t^a_{34} \,2\sqrt{x_3 x_4} \,\, \{\} \, , 
\end{eqnarray}
where the brackets $\{\}$ contain,
\begin{eqnarray}
 \{\}\, & = & \, \varepsilon_1^*\varepsilon_2^* \, 
  {x_2 - x_1 \over 2(x_1+ x_2)^2 } \, \left[
  \tilde r_{1+2,1} \tilde r_{3+4,3}\,
  \xi_{-3}^\dagger\sigma^3\chi_4\,
  a_1^\dagger a_2^\dagger d_3 b_4  
  +
  \tilde r_{1+2,1} \tilde r_{4,3}\,
  \chi_{3}^\dagger\chi_4\,
  b_3^\dagger a_1^\dagger a_2^\dagger b_4  
  -
  \tilde r_{1+2,1} \tilde r_{3,4}\,
  \xi_{-3}^\dagger\xi_{-4}\,
  d_4^\dagger a_1^\dagger a_2^\dagger d_3
  \right] \nonumber \\ 
& - & 
  \varepsilon_1^*\varepsilon_2 \,  
  {x_1 + x_2 \over (x_3 + x_4)^2}\,
  \tilde r_{3+4,3} \tilde r_{2,1}\,
  \chi_3^\dagger\sigma^3\xi_{-4}\,
  b_3^\dagger d_4^\dagger a_1^\dagger a_2 \nonumber \\
& + &
 \varepsilon_1^*\varepsilon_2 \,  
  {x_1 + x_2 \over (x_1 - x_2)^2}\,
  \left[
  \left(\tilde r_{2,1} \tilde r_{4,3}+ \tilde r_{1,2} \tilde r_{3,4}\right)\,
  \xi_{-3}^\dagger\xi_{-4}\,
  d_4^\dagger a_1^\dagger a_2 d_3 
  -  
  \left(\tilde r_{2,1} \tilde r_{3,4}+ \tilde r_{1,2} \tilde r_{4,3}\right)\,
  \chi_3^\dagger\chi_4\,
  b_3^\dagger a_1^\dagger a_2 b_4 
  \right] + h.c. \, .
\end{eqnarray}
Finally, the instantaneous gluon interaction between quarks reads
\begin{eqnarray}
  H_{(\psi\psi)^2} & = & \sum_{1234}\int[1234] 
  \tilde\delta(p^\dagger - p)\, {g^2 \over 2}\, 4\sqrt{x_1 x_2 x_3 x_4} \,\,\{\} \, , 
\end{eqnarray}
where the brackets $\{\}$ contain,
\begin{eqnarray}
\{\} & = &
- {1 \over 2} \left[
  {\chi_1^\dagger \chi_2 \chi_3^\dagger \chi_4 \over (x_1-x_2)^2}\, 
  t^a_{12} t^a_{34}\, 
  [\tilde r_{1,2} \tilde r_{4,3}+ \tilde r_{2,1} \tilde r_{3,4}]
  -{\chi_3^\dagger \chi_2 \chi_1^\dagger \chi_4 \over (x_3-x_2)^2}\, 
  t^a_{32} t^a_{14}\,
  [\tilde r_{3,2} \tilde r_{4,1}+ \tilde r_{2,3} \tilde r_{1,4}]\,
  \right]\,
  b_1^\dagger b_3^\dagger b_2 b_4  \nonumber \\ 
& + &
  {1 \over 2} \left[
  {\xi_{-2}^\dagger \xi_{-1} \xi_{-4}^\dagger \xi_{-3} \over (x_1-x_2)^2}\,
  t^a_{21} t^a_{43}\,
  [\tilde r_{1,2} \tilde r_{4,3}+ \tilde r_{2,1} \tilde r_{3,4}]\, 
  -{\xi_{-2}^\dagger \xi_{-3} \xi_{-4}^\dagger \xi_{-1} \over (x_3-x_2)^2}\,
  t^a_{23} t^a_{41}\,
  [\tilde r_{3,2} \tilde r_{4,1}+ \tilde r_{2,3} \tilde r_{1,4}]\,
  \right]\,
  d_1^\dagger d_3^\dagger d_2 d_4 \nonumber \\
& + &
  \left(
  \left[
  {\chi_1^\dagger \chi_2 \chi_3^\dagger \sigma^3 \xi_{-4} \over (x_1-x_2)^2}\,
  t^a_{12} t^a_{34}\,
  \tilde r_{2,1} \tilde r_{3+4,3}\,
  -{\chi_3^\dagger \chi_2 \chi_1^\dagger \sigma^3 \xi_{-4} \over (x_3-x_2)^2}\,
  t^a_{32} t^a_{14}\,
  \tilde r_{2,3} \tilde r_{1+4,1}\,
  \right]\,
  b_1^\dagger b_3^\dagger d_4^\dagger b_2 + h.c. \right) \nonumber \\
& - &
  \left(
  \left[
  {\chi_1^\dagger\sigma^3\xi_{-2}\xi_{-4}^\dagger\xi_{-3}\over(x_1+x_2)^2}\,
  t^a_{12} t^a_{43}\,
  \tilde r_{4,3} \tilde r_{1+2,1}\,
  -{\chi_1^\dagger\sigma^3\xi_{-3}\xi_{-4}^\dagger\xi_{-2}\over(x_1+x_3)^2}\,
  t^a_{13} t^a_{42}\,
  \tilde r_{4,2} \tilde r_{1+3,1}\,
  \right]\,
  b_1^\dagger d_2^\dagger d_3^\dagger d_4 + h.c. \right) \nonumber \\
& - & 
  2\, 
  {\chi_1^\dagger \chi_2 \xi_{-4}^\dagger \xi_{-3} \over (x_1-x_2)^2}
  t^a_{12} t^a_{43}
  [\tilde r_{1,2} \tilde r_{4,3}+ \tilde r_{2,1} \tilde r_{3,4}]\, 
  b_1^\dagger d_3^\dagger d_4 b_2 + 
  2\, 
  {\chi_1^\dagger\sigma^3\xi_{-3}\xi_{-4}^\dagger\sigma^3\chi_2 
  \over (x_1+x_3)^2}\,
  t^a_{13} t^a_{42}\,
  \tilde r_{1+3,1} \tilde r_{2+4,2}\,
  b_1^\dagger d_3^\dagger d_4 b_2 \, .
\end{eqnarray}
\end{widetext}
Useful color identities are:
$ t^at^bt^a = {- t^a/(2N_c)}$, $t^at^b+t^bt^a = {\delta^{ab}/N_c}
+ d^{abc}t^c$, $d^{abc}d^{abd} = [(N_c^2 - 1)/N_c]\delta^{cd}$, 
$f^{abc}t^bt^c = i (N_c/2)t^a$. 

\section{$Q_\lambda \bar Q_\lambda$ interaction}
\label{A:Q}
Several factors are needed to estimate the small-$x$ behavior of the 
potential kernel $\tilde v_0(13,24)$ in Eq. (\ref{v0}). Momenta are 
labeled according to Figs. \ref{fig:oge} and \ref{fig:sea}.
\begin{equation}
f_{13,24} = \exp{\left[ - ({\cal M}^2_{13} - 
{\cal M}^2_{24})^2/\lambda^4\right]} \, ,
\end{equation}
where 
\begin{eqnarray}
\label{kij}
{\cal M}^2_{ij} & = & 4(m^2 + |{\vec k}_{ij}|^2) \nonumber \\
& = &  { \kappa_{ij}^{\perp \, 2} + m^2 \over x_i x_i }  \, ,
\end{eqnarray}
with 
\begin{eqnarray}
k_{ij}^\perp = \kappa_{ij}^\perp \, , \\
k_{ij}^3            = (x_i-1/2){\cal M}_{ij} \, .
\end{eqnarray}
\begin{eqnarray}
{\cal M}^2_{13} - {\cal M}^2_{24} = 4 ({\vec k}_{13} + {\vec k}_{24})\, \vec q \, ,
\end{eqnarray}
where
\begin{eqnarray}
\label{vecq}
\vec q = {\vec k}_{13} - {\vec k}_{24} \, ,
\end{eqnarray}
is the momentum transfer that goes over to the standard one in the NR limit. 
\begin{eqnarray}
\label{13-24}
{\cal M}^2_{13} - {\cal M}^2_{24} =  
{ 2 \kappa^\perp_{13} q^\perp  - q^{\perp \, 2}  - z(1-2x_1 + z) {\cal M}_{13}^2
\over (x_1-z)(x_3+z) } \, . \nonumber \\
& &
\end{eqnarray}
In the last term in Eq. (\ref{v0}), the form factors $f_{1,52}f_{53,4}$
have arguments 
\begin{equation}
\label{m52}
{\cal M}^2_{52} - m^2 = { x_1 \over x_1 - z } \, {\cal D}_1 \, ,
\end{equation}
where 
\begin{equation}
{\cal D}_1  \equiv  {x_1 \over |z| } \, 
\left(q^\perp - {z \over x_1}  \kappa^\perp_{13} \right)^2
+ m^2 { |z| \over x_1 } \, ,
\end{equation}
and, 
\begin{eqnarray}
\label{m53}
{\cal M}^2_{53} - m^2 & = & {\cal D}_3  \, ,
\end{eqnarray}
where
\begin{eqnarray}
{\cal D}_3 = {x_3 \over |z|} \, 
\left(q^\perp + {z \over x_3} \kappa^\perp_{13} \right)^2
+ m^2 { |z| \over x_3} \, ,
\end{eqnarray}
while the form factors $f_{3,54}f_{51,2}$ have the arguments,
\begin{equation}
\label{m54}
{\cal M}^2_{54} - m^2 = { x_3 \over x_3 + z}\, {\cal D}_3 \, ,
\end{equation}
and, 
\begin{equation}
\label{m51}
{\cal M}^2_{51} - m^2 = {\cal D}_1 \, .
\end{equation}
The last term in Eq. (\ref{v0}) can be written with the coefficient $1 = 
(1 - f_{13,24}) + f_{13,24}$ but only the second term counts at small $z$
because $(1 - f_{13,24})$ is proportional to the momentum transfer squared.
The factor  $f_{13,24}$ becomes common to all terms in Eq. (\ref{v0}) and 
is taken out in front. The LF instantaneous term can be split into the part 
$ff$ that joins the low-energy exchange $1-ff$ that goes with the high-energy 
exchange. This way one obtains 
\begin{widetext}
\begin{eqnarray}
\label{evsmallx}
\left({\kappa^{\perp \, 2}_{13} + m^2_{0} \over x_1 x_3} +
{\tilde m^2_{1}\over x_1} + {\tilde m^2_{3}\over x_3}  - M^2 \right)
  \tilde \psi(\kappa^\perp_{13},x_1) 
  - {4 \alpha \over 3\pi^2 } \int dx_2 d^2 \kappa^\perp_{24} 
  \, \sqrt{x_1 x_3 \over x_2 x_4} \, { f_{13,24} \over x^2_5 }
  \, \tilde v_0(13,24) \, \tilde \psi(\kappa^\perp_{24},x_2) = 0 \, ,
\end{eqnarray}  
where 
\begin{eqnarray}
\label{vsplit}
\tilde v_0(13,24)  & = &  
\theta(z)  \tilde r_{5/1} \tilde r_{5/4} \, 
(\tilde v_{+\, high} + \tilde v_{+\, low})
+ 
\theta(-z) \tilde r_{5/3} \tilde r_{5/2} \, 
(\tilde v_{-\, high} + \tilde v_{-\, low}) \, .
\end{eqnarray}
The gluon mass ansatz contributes to the low-energy exchange terms 
only. In terms of the invariant masses from Eqs. (\ref{m52} - \ref{m51}),
\begin{eqnarray}
\label{vh+}
\tilde v_{+\, high}  & = &  
 { f_{1,52} f_{53,4} - 1 \over 2 } \, \left\{
  {x_1^2 + x_4^2 \over x_1 x_4 } \,
{({\cal M}^2_{52} - m^2 )({\cal M}^2_{53} - m^2)
  \over  ({\cal M}^2_{52} - m^2 )^2 + ({\cal M}^2_{53} - m^2)^2 } 
       -  1 \right\} \, ,
\end{eqnarray}
\begin{eqnarray}
\label{vh-}
\tilde v_{-\, high} & = &  
 { f_{3,54} f_{51,2} - 1 \over 2} \,  \left\{
  {x_3^2 + x_2^2 \over x_3 x_2 } \,
{({\cal M}^2_{54} - m^2 )({\cal M}^2_{51} - m^2)
  \over  ({\cal M}^2_{54} - m^2 )^2 + ({\cal M}^2_{51} - m^2)^2 } 
       -  1 \right\} \, .
\end{eqnarray}
These have the same limit when $z \rightarrow 0$ for fixed $q^\perp$,
\begin{eqnarray}
\lim_{z \rightarrow 0} \tilde v_{+\, high} & = &
\lim_{z \rightarrow 0} \tilde v_{-\, high} \nonumber \\
& = &
 { f_{3,54} f_{51,2} - 1 \over 2} \,  
 {x_1 - x_3 \over x_1 x_3 (x_1^2 + x_3^2) } \,
 {(q^\perp - \kappa^\perp_{13})^2 - \kappa^{\perp \, 2}_{13}
  \over q^{\perp \, 2}  } \,\, z  \, + \, o(z^2) \, .
\end{eqnarray}
The terms on the order of $z^2$ and higher are finite when
divided by the square of $x_5 = |z|$. Terms linear in $z$ 
produce an integral convergent in the sense of principal
value \cite{z, long}. When $q^\perp \sim \sqrt{z} \rightarrow 0$, 
$d^2 \kappa^\perp_{24}$ removes one power of $z$ from the denominator 
in Eq. (\ref{ev0}), while $\tilde v_{\pm \, high}$ vanish for 
$z \rightarrow 0$. The contributions of $\tilde v_{\pm \, high}$ 
are $\vec k^{\,2}/m^2$ times smaller than the dominant terms
and can be ignored in the first approximation. One can see this 
by integrating $\tilde v_{\pm\, high}$ with a Coulomb wave function.

The low-energy terms read
\begin{eqnarray}
\label{v+low}
\tilde v_{+\, low} & = & 
{f_{1,52} f_{53,4} \over 4} \,
\left[ 2
- {  ({\cal M}^2_{53} - m^2 )/x_4 - \mu^2(2,5,3)/x_5 \over  
     ({\cal M}^2_{52} - m^2 )/x_1 + \mu^2(2,5,3)/x_5      }  
- {  ({\cal M}^2_{52} - m^2 )/x_1 - \mu^2(2,5,3)/x_5 \over  
     ({\cal M}^2_{53} - m^2 )/x_4 + \mu^2(2,5,3)/x_5      }   
\right] \, \, , \nonumber  \\
& & 
\end{eqnarray}
\begin{eqnarray}
\label{v-low}
\tilde v_{-\, low}  & = & 
{f_{3,54} f_{51,2} \over 4} \,
\left[ 2
- {  ({\cal M}^2_{51} - m^2 )/x_2 - \mu^2(1,5,4)/x_5 \over  
     ({\cal M}^2_{54} - m^2 )/x_3 + \mu^2(1,5,4)/x_5      }   
- {  ({\cal M}^2_{54} - m^2 )/x_3 - \mu^2(1,5,4)/x_5 \over  
     ({\cal M}^2_{51} - m^2 )/x_2 + \mu^2(1,5,4)/x_5      }   
\right] \, . \nonumber \\
& &
\end{eqnarray}
\end{widetext}
\section{ Mass terms }
\label{A:M}
The mass terms with $i=1, 3$ in the eigenvalue Eq. (\ref{evsmallx}) are
given in Eq. (\ref{mstildemi}), with ${\cal M}^2_1$ given in Eq. (\ref{m1}), 
and ${\cal M}^2_3$ in Eq. (\ref{m3}). $m^2_0$ originates from Eq. (\ref
{mlambda}) with $\lambda = \lambda_0$. Namely, the quark mass counterterm 
in $X$ of Eq. (\ref{H}) adds $\delta m^2_{\Delta \delta}$ to the original 
mass parameter $m^2$ in Eq. (\ref{tquark}) and the free ultraviolet-finite 
part of the counterterm is such that $m^2_0$ appears in Eq. (\ref{ev0}).  
The condition on $m_0^2$ that the eigenstates of $H_{\lambda_0}$ with quantum 
numbers of a single quark have eigenvalues growing to infinity is fulfilled 
below by representing gluon interactions in the case of the single quark state 
by a new gluon mass ansatz. The resulting value of $m_0^2$ enters into the 
quarkonium dynamics. The determination of the ultraviolet-finite part of the 
mass counterterm in $X$ in Eq. (\ref{H}) is thus based on the picture that 
comes out from simultaneous consideration of two eigenvalue equations, one for 
the state with quantum numbers of a single quark (or an anti-quark, the result 
is the same), and another one for the quarkonium. The key physical assumption 
made in the comparison is that the binding of effective quarks in the quarkonium 
state occurs at the expense of change in their individual structure. While the 
buildup of self-interacting clouds of gluons around single quarks leads to the 
infinite quark masses, in the case of a colorless pair the main parts of the 
gluon clouds can recombine into a colorless object that may fly out of the 
region of strong interaction with the quarks, leaving behind only the minimal 
remnants of the gluon clouds required to form the quarkonium eigenstate with 
a finite mass. The new finite balance is described using the gluon mass ansatz 
parameter $\delta_\mu$. The finite balance can be achieved because the 
quark-anti-quark state looks neutral from large distances and does not continue 
to generate gluons over infinite distances along the LF. This scenario is 
partly similar to the one originally developed in the LF dynamics in \cite{gcoh2, 
gcoh3}, and studied in \cite{QQgcoh, QQ}. The main differences are related to 
the fact that the physical picture that emerges here in the finite effective 
theory with the gluon mass ansatz relies on the phenomenological parameter 
$\delta_\mu$. A formal cutoff parameter $\delta^+$ of the canonical theory, 
the coupling coherence phenomenon that may work over many scales of an ultraviolet 
cutoff, and the condition of transverse locality are not employed in the new 
picture. Instead, the present scenario can be studied in higher orders of 
perturbation theory according to the known rules \cite{g_QCD, optimization1} 
that explicitly preserve the boost invariance, cluster decomposition property, 
and unitary connection with the initial theory. 

The eigenstate of $H_{\lambda_0}$ with a single quark quantum numbers
and momentum $p$ with components $p^+$ and $p^\perp$, has the eigenvalue 
\begin{equation}
\label{qe}
p^- = (p^{\perp \, 2} + \tilde m^2)/p^+ \, ,
\end{equation} 
and the decomposition in the effective particle basis,
\begin{equation}
\label{|p>}
|p\rangle = |Q_{\lambda_0} \rangle + |Q_{\lambda_0} g_{\lambda_0} \rangle + 
            \, . \, . \, . \, .
\end{equation}
The new gluon mass ansatz is introduced in the quark-gluon component. 
It is different than in the quarkonium case because the states have 
different quantum numbers and are made of different numbers of effective 
particles. Dropping the subscript $\lambda_0$ as in Eq. 
(\ref{matrix}), the eigenvalue problem is written as
\begin{eqnarray}
\label{Qmatrix}
(T_q + \tilde T_g)\,|Qg\rangle \, + \, Y|Q\rangle 
& = & E\,|Qg\rangle \, ,\nonumber \\ 
Y\,|Qg\rangle \, + \, 
T_q\,|Q\rangle & = & E\,|Q\rangle  \, . 
\end{eqnarray}
The new ansatz enters through the kinetic energy $\tilde T_g$, which contains
\begin{eqnarray}
\label{mul1q}
\tilde \mu^2_{\lambda_0} = \mu^2_{\lambda_0} + \mu_Q^2(x, \kappa^\perp) \, ,
\end{eqnarray}
where $x$ and $\kappa^\perp$ refer to the relative motion of
the effective gluon with respect to the quark. The operator 
$R$ from Eq. (\ref{R}) is now replaced by the one with $\hat 
{\cal P}$ that projects on the single effective quark basis 
state with kinematical momentum components $p^+$ and $p^\perp$.
In the perturbative expansion in $g$, only the second term 
on the right-hand side of Eq. (\ref{mul1q}) contributes in
orders up to $g^2$. Thanks to the boost invariance, the resulting 
eigenvalue condition reduces to an equation for $\tilde m^2$, 
which is independent of $p$,
\begin{eqnarray}
\label{evq}
\tilde m^2 & = & 
m^2_{0} - {4 \alpha_0 \over 3 \pi^2} \, 
\int dx d^2 \kappa^\perp \, \tilde r^2_\delta(x)\, 
\nonumber \\
& \times & f^2_{\lambda_0} (m^2,{\cal M}^2)\,  
{1\over x^2}\, 
{ {\cal M}^2 - m^2 \over {\cal M}_Q^2 - m^2} \, .
\end{eqnarray}
\begin{eqnarray}
{\cal M}_Q^2 = [\kappa^{\perp \, 2}+ \mu_Q^2(x, \kappa^\perp)]/x
     + (\kappa^{\perp \, 2}+ m^2)/(1-x) \, . \nonumber \\
\end{eqnarray}
For $\tilde m^2$ to be positive and growing to infinity when 
$\delta \rightarrow 0$, one can write $m_{0}^2$ in the 
integral form,
\begin{eqnarray} 
\label{resultmu0phi}
m^2_{0} &=& m^2 + {4 \alpha_0 \over 3 \pi^2} \, 
\int dx d^2 \kappa^\perp \, \tilde r^2_\delta(x)\, 
\nonumber \\
& \times & f^2_{\lambda_0} (m^2,{\cal M}^2)\,  
{1\over x^2}\, 
{ {\cal M}^2 - m^2 \over {\cal M}_{0}^2 - m^2} \, ,
\end{eqnarray}
with some function ${\cal M}_{0}^2$ that satisfies the condition,
\begin{equation}
{\cal M}_Q^2 > {\cal M}_{0}^2 > m^2 \, .
\end{equation}
This condition can be satisfied by writing, 
\begin{eqnarray}
\label{M0phi}
{\cal M}_{0}^2 = [\kappa^{\perp \, 2}+ \mu_{0}^2(x, \kappa^\perp)]/x
     + (\kappa^{\perp \, 2}+ m^2)/(1-x) \, , \nonumber \\
\end{eqnarray}
and assuming that
\begin{equation}
\label{massansatzinequality}
\mu_Q^2 > \mu_{0}^2 \ge 0 \, .
\end{equation}
As long as the difference $\mu_Q^2 - \mu_{0}^2$ does not 
vanish for $x \rightarrow 0$, the single quark mass will
tend to $\infty$ when $\delta \rightarrow 0$. But this may 
easily happen here because the larger is the gluon mass ansatz 
$\mu_Q^2$, the stronger the single quark mass eigenvalue 
diverges in the limit $\delta \rightarrow 0$, while $\mu_{0}^2$ 
remains free to vanish in the limit $x \rightarrow 0$ and lead
to a finite mass contribution in the quarkonium dynamics. Using 
Eq. (\ref{resultmu0phi}) for $m_0^2$ one obtains Eq. (\ref{mstildemi}).

\end{document}